\documentclass[12pt, draftclsnofoot, onecolumn]{IEEEtran}

\input epsf
\usepackage{graphicx,epstopdf}   
\usepackage{pifont,dsfont}
\usepackage{bm,bbm}
\usepackage{float}
\usepackage[utf8x]{inputenc}
\usepackage{color}
\usepackage{mathtools}
\usepackage[table]{xcolor}
\usepackage{subfigure}

\definecolor{Gray}{gray}{0.9}

\usepackage[linesnumbered,algoruled,boxed,lined]{algorithm2e} 
\usepackage{amssymb,amsmath,amsfonts,amsthm}
\usepackage{cite,citesort}
\usepackage{balance}
\usepackage[utf8x]{inputenc}
\usepackage{graphicx}
\usepackage{epsfig}

\usepackage[
top    = 1.70cm,
bottom = 1.05in,
left   = 0.60 in,
right  = 0.60 in]{geometry}

\IEEEoverridecommandlockouts

\newtheorem{theorem}{Theorem}

\newtheorem{corollary}{Corollary}

\begin{document}

\title{A Decoupled Learning Strategy for Massive Access Optimization in Cellular IoT Networks}

\author{
\IEEEauthorblockN{Nan Jiang, Yansha Deng, Arumugam Nallanathan, and Jinghong Yuan}\\

\thanks{N. Jiang, and A. Nallanathan are with School of Electronic Engineering and Computer Science, Queen Mary University of London, London, UK (e-mail: \{nan.jiang, a.nallanathan\}@qmul.ac.uk).
}
\thanks{Y. Deng is with Department of Informatics, King's College London, London, UK (e-mail: yansha.deng @kcl.ac.uk). 
}
\thanks{J. Yuan is with School of Electrical Engineering and Telecommunications, University of New South Wales, Sydney, Australia (e-mail: j.yuan@unsw.edu.au). 
}

\vspace*{-1.6cm}
}



\maketitle

\begin{abstract}

Cellular-based networks are expected to offer connectivity for massive Internet of Things (mIoT) systems. However, their Random Access CHannel (RACH) procedure suffers from unreliability, due to the collision from the simultaneous massive access. Despite that this collision problem has been treated in existing RACH schemes, these schemes usually organizes IoT devices' transmission and re-transmission along with fixed parameters, thus can hardly adapt to time-varying traffic patterns. Without adaptation, the RACH procedure easily suffers from high access delay, high energy consumption, or even access unavailability. With the goal of improving the RACH procedure, this paper targets to optimize the RACH procedure in real-time by maximizing a long-term hybrid multi-objective function, which consists of the number of access success devices, the average energy consumption, and the average access delay. To do so, we first optimize the long-term objective in the number of access success devices by using Deep Reinforcement Learning (DRL) algorithms for different RACH schemes, including Access Class Barring (ACB), Back-Off (BO), and Distributed Queuing (DQ). The converging capability and efficiency of different DRL algorithms including Policy Gradient (PG), Actor-Critic (AC), Deep Q-Network (DQN), and Deep Deterministic Policy Gradients (DDPG) are compared. Inspired by the results from this comparison, a decoupled learning strategy is developed to jointly and dynamically adapt the access control factors of those three access schemes. This decoupled strategy first leverage a Recurrent Neural Network (RNN) model to predict the real-time traffic values of the network environment, and then uses multiple DRL agents to cooperatively configure parameters of each RACH scheme. Specifically, each RACH scheme is optimized using its preferred DRL algorithms, where DQN is applied to handle discrete action selections (BO and DQ), and DDPG is applied to handle continuous action selection (ACB). Numerical results showcase that our proposed decoupled learning strategy speeds up the training about 10 times than the conventional strategy, and it remarkably outperforms conventional strategy after converging.


\end{abstract}

\section{Introduction}

Cellular-based radio access technologies are required to support massive Internet of Things (mIoT) ecosystem, due to its high reliability, security, and scalability. In most state-of-the-art IoT systems, including enhanced Machine-Type Communication (eMTC), NarrowBand (NB)-IoT, and 5G New Radio (NR), Random Access CHannel (RACH) procedures have been adopted to establish synchronization between IoT devices and Base Stations (BSs). In these systems, time is organized into frames (a.k.a transmission time interval), where each frame consisting of multiple preambles (a.k.a. RACH channels) for IoT devices to request access. In each frame, IoT devices are activated by the requests from their upper layer applications that is unknown to the BS, in other words, these devices try to connect to the associated BS in an uncoordinated manner by selecting preambles at random. Due to this uncoordination, collisions occur when IoT devices select the same preamble at the same frame, which inevitably increases the access delay and the energy consumption, or even leads to service unavailability. To solve this problem, several access control schemes have been proposed in traditional access control works, including Access Class Baring (ACB) \cite{wu2012fast,duan2016d,he2017traffic}, Back-Off (BO) \cite{jian2013novel}, Distributed Queuing (DQ) \cite{vazquez2017contention}, and etc.. These schemes aim to improve the access success probability of RACH by intelligently organizing preamble transmission and re-transmission of IoT devices, but, its efficiency strongly depends on the incoming traffic of the IoT devices, which is generally intractable and dynamic over time.

With the goal of improving the access success, majority of efforts in previous works \cite{wu2012fast,duan2016d,he2017traffic,jian2013novel,vazquez2017contention,Luis2018Reinforcement} have been devoted to formulate a mathematical model to describe the regularities of the practical communication environment as well as the traffic access pattern, so as to explicitly optimize the access control factor of each RACH scheme. Classical dynamic ACB optimizations have been studied in \cite{wu2012fast,duan2016d,he2017traffic}, where the ACB factor was configured based on the approximated future backlog estimation conducted by the methods of drift analysis \cite{wu2012fast}, Method of Moments (MoM) \cite{duan2016d}, and Maximum-Likelihood Estimation (MLE) \cite{he2017traffic}, respectively. In \cite{jian2013novel}, the class-dependent BO scheme has been proposed to guarantee the acceptable access delay of IoT devices with different priorities. In \cite{vazquez2017contention}, a DQ scheme based on tree-splitting algorithm has been proposed to perfectly solve collisions, but how to select the tree-splitting factor for access optimization has not been discussed due to its analytical intractability. More recently, to fulfill mission-critical IoT applications, the network is expected to provide not only reliable wireless access, but also the low access delay and energy consumption. In \cite{bello2018energy,azari2019latency,wali2018optimization}, the energy-delay tradeoff in the RACH optimization has been studied from the perspectives of optimized extended ACB \cite{bello2018energy}, power saving mode \cite{azari2019latency}, and repetition values \cite{wali2018optimization}, respectively. However, due to its mathematical complexity, all these works balance the energy-delay tradeoff using queuing frameworks based on a static analytical model by ignoring the dynamic of the network system.

To deal with more complex communication environment and practical formulations, Machine Learning (ML), specifically Reinforcement Learning (RL), emerges as a promising tool to optimize RACH, due to that it solely relies on the self-learning of the environment interaction, without the need to derive explicit optimization solutions based on a complex mathematical model. Recent work \cite{Luis2018Reinforcement} optimized the ACB scheme based on a tabular Q-learning algorithm, but this tabular method can not be used to solve other access optimization problems, due to its inefficiency in handling large state and action space. The most relevant work is our prior work \cite{jiang2019cooperative,jiang2019deep}, where we developed a cooperative DQN algorithm for uplink resource configuration to optimize the number of served IoT devices in NB-IoT networks. However, these learning-based RACH optimizations only focused on maximizing the number of access success devices at the BS side, whereas ignored the delay and energy consideration at the user side. Knowing the importance of the energy-delay tradeoff for mission critical IoTs \cite{bello2018energy,azari2019latency,wali2018optimization}, there is a need for an dynamically optimized RACH satisfying the access, energy, and delay requirements simultaneously in order to support massive access with delay and energy consideration in industrial IoT scenario.

To solve this problem, in this paper, we aim to develop a novel learning strategy to efficiently optimize the hybrid performance metric, taking into account the number of success accesses, the energy consumption of IoT devices, and the access delay of IoT devices for three main access control schemes. Our main contributions can be summarized as follows:
\begin{itemize}
\item 
To effectively optimize the existing RACH schemes, we first propose four DRL algorithms, including Policy Gradient (PG), Actor-Critic (AC), Deep Q-Network (DQN), and Deep Deterministic Policy Gradients (DDPG), where the PG, AC, and DQN target to optimize the BO and the DQ schemes with discrete action space, and the DDPG aims at optimizing the ACB scheme with continuous action space. All the DRL algorithms leverage the Recurrent Nerual Network (RNN) model, specifically, the Gated Recurrent Unite (GRU) architecture, to approximate their value function/policy. Implementing RNN model not only enables learning operation over a large state space, but also help to capture the complex time correlation among each transmission frame that would help the DRL agent to better maximize the future performance. Our results shown that our proposed DRL-based RACH schemes significantly outperform conventional heuristic schemes in terms of the number of access success devices.
\item 
We then present a novel method to balance the energy-delay tradeoff in an online manner, where each RACH scheme is jointly optimized in terms of a hybrid Key Performance Indicators (KPIs) objective, including the number of access success, the energy consumption, and the access delay. The importance of each KPI can be adapted by configuring their weights, so as to fulfil different performance requirements of the network. To handle numerous action space conducted by multiple RACH schemes, we leverage the cooperative multi-agent DRL, where each DRL agent independently handles a single scheme, and shares their historical action selections to enable cooperation. The proposed hybrid scheme outperforms other single schemes in most combinations of KPIs. 
\item 
In order to efficiently train these parallel DRL agents, we propose a novel learning strategy, where an RNN predictor is used to predict overall traffic statistics, and such a predicted value is inputted into each DRL agent as the belief state to configure parameter for each RACH scheme. The proposed learning strategy is able to be updated in an online manner, thus it is possible to be pre-trained in the simulation environment, and then to be transferred to practice. The proposed decoupled learning strategy can achieve better performance than the conventional strategies, and uses much less training time. 
\item 
Finally, our proposed decoupled learning strategy is applied to optimize the number of served IoT devices in a practical case, namely, multi-group NB-IoT network. The simulation environment takes into account the RACH as well as the uplink channel resource scheduling based on the 3GPP reports \cite{3GPP2015Cellular,3GPP2017MAC}. The result shown that our proposed decoupled learning strategy speeds up the training about 10 times than the conventional schemes.

\end{itemize}

The rest of the paper is organized as follows. Section II formulates the problem and illustrates the system model. Section III provides preliminary and DRL-based single RACH scheme optimization. Section IV presents the hybrid scheme optimization and the decoupled learning strategy. Finally, Section V concludes the paper.

\vspace*{-0.3cm}
\section{System Model and Problem Formulation}\label{sec3}
\vspace*{-0.1cm}

We consider a cellular-based mIoT network system consisting of an arbitrary number of IoT devices and a single BS, where they are in-synchronized, thus are unaware of the status of each other. For each IoT device, there is an random process for the generation of uplink data packets, which is unknown to the BS. We consider the time is divided into frames, and the packets inter-arrival processes are independent and identically distributed over the time frames, which are Markovian as defined in \cite{3GPP2011Study,3GPP2015Cellular}.

\vspace*{-0.3cm}
\subsection{Problem Formulation}\label{sec3.1}
\vspace*{-0.1cm}

We model the grant-based RACH procedure, where every IoT device has only two possible states\footnote{The \textit{grant-based} RACH procedure separates random access and grant-based data transmission, where packets in an IoT device will be completely transmitted if it access succeeds (assuming there are enough uplink channel resource). The \textit{grant-free} access integrates access preamble and data into one sequence, where multiple sequences might be accumulated in the queue of an IoT device waiting for transmission. The proposed model in this paper can also support the study of \textit{grant-free} access, with the minor modification by considering multiple packets queuing.}, either \textit{inactive} or \textit{active}, while an IoT device with uplink data packets to be transmitted is in the latter case. Once active, an IoT device executes the RACH procedure in order to establish the synchronization with the BS. Without loss of generality, we focus on the RACH, and assume that the generated uplink packets by an IoT device would be completely transmitted if it succeeds in access. This assumption was specified in 3GPP report \cite{3GPP2011Study}, and was considered in several prior works \cite{he2017traffic,duan2016d,wu2012fast,jiang2017random,jiang2018rach,jiang2018collision,jiang2019deep}.

The RACH procedure consists of four steps. The first step, based on framed-ALOHA principle, allows an IoT device to transmit a randomly selected preamble. The latter three steps, based on BS's scheduled channel, use for control information exchanges \cite{LTE2013dahlman} (details to be discussed in Sec. \ref{sec3.2}). The RACH can fail if a collision occurs between two or among more IoT devices selecting the same preamble in Step 1. The collision can frequently occur when massive IoT devices try to access simultaneously, which increases the access latency and the energy consumption. To handle this, one can allocate the transmission and re-transmission of these IoT devices using access control schemes to spread the traffic loads. More specifically, we focus on the most common access control schemes, which are the ACB, the BO, and the DQ schemes. The ACB scheme, forbidding IoT devices to access according to a dynamic probability $\beta_\text{ACB}$, can effectively alleviate network congestion, but always results in severe energy consumption when accesses are overly forbidden. The BO scheme, requiring IoT devices to postpone their access attempts for a period within $[2^{\beta_\text{BO}-1}, 2^{\beta_\text{BO}}]$ frames when collision occurs, considers to both congestion alleviation and energy saving, but always leads to a large access delay. The DQ scheme, organizing the transmission and re-transmission using a tree splitting rule with dynamic depth $\beta_\text{Tdepth}$ and degree $\beta_\text{Tdegree}$, is deem to intelligently resolve the collisions with minimized energy consumption and delay, but is hard to be globally optimized over time due to its complexity.

In this work, we aim to tackle the problem of optimizing the access control factors defined as $A^t=\{\beta^t_\text{ACB}, \beta^t_\text{BO}, \beta^t_\text{Tdepth}, \beta^t_\text{Tdegree}\}$ in an online manner for every frames. At the beginning of each frame $t$, the decision is made by the BS according to the transmission receptions $U^{t'}$ for all prior frames $t'=1,...,t-1$, which consists of the following variables: the number of the access success devices $V^{t'}_{\rm s}$, the number of the collided preambles $V^{t'}_{\rm c}$, the number of idle preambles $V^{t'}_{\rm i}$, the average energy consumption of each success device $V^{t'}_{\rm e}$, and the average access delay of each success device $V^{t'}_{\rm d}$. Without loss of generality, we assume that each device $j$ would report their energy consumption ${E}_j$ and access delay ${D}_j$ to the BS if their access succeeded. The access delay ${D}_j$ of device $j$ is normalized by counting the number of frames that this device was used for RACH, while the energy consumption ${E}_j$ summarizes the energy that the device $j$ was used for receiving system information and executing RACH (to be detailed in Sec. \ref{sec3.2.2}). Note that the observation in each frame $t$ includes all histories of such measurements and past actions, which is denoted as $H^{t}$=$\{O^{1},O^{2},...,O^{t-1}\}$, each with $O^{t-1}=\{U^{t-1},A^{t-1}\}$.

At each frame $t$, the BS aims at maximizing a long-term objective ${R}^t$ (reward) related to the average access success es $V_{\rm s}^t$, the average energy consumption $V_{\rm e}^t$, and the average access delay $V_{\rm d}^t$. The optimization relies on the selection of parameters in $A^t$ according to the current historical observation $O^{t}$ with respect to the stochastic policy $\pi$. This optimization problem can be formulated as:
\vspace*{-0.2cm}
\begin{align}\label{eq1}
&(\text{P1}):  \mathop {\text{max}}\limits_{ \pi(A^t|O^t)}     \quad   \sum_{k=t}^{\infty}  \gamma^{k-t} {R}^k,
\end{align}
where $\gamma \in [0,1)$ is the discount factor for the performance accrued in future frames, and the objective function ${R}^t$ is formulated as 
\vspace*{-0.2cm}
\begin{align}\label{eq2}
{R}^t  = x_{s} V_{\rm s}^t +  x_{d} {R_{\rm d}^t} + x_{e} {R_{\rm e}^t}.
\end{align}
In \eqref{eq2}, $x_{s}$, $x_{d}$ and $x_{e}$ are the weights of the success accesses reward $R_{\rm s}^t$, the access delay reward ${R_{\rm d}^t}$, and the energy consumption reward ${R_{\rm e}^t}$, respectively. Note that these three sub rewards are obtained by normalizing the observation of the average success accesses $V_{\rm s}^t$, the average access delay reward ${V_{\rm d}^t}$ of each succeeded device, and the average energy consumption ${V_{\rm e}^t}$ of each succeeded device, respectively. The using of weighted rewards are inspired by the study of multi-criteria RL \cite{gabor1998multi,Sunehag2018}, which targets to solve the problem that typically without available trade-offs among different objectives a priori. The derivation of these three sub rewards will be detailed in Sec. \ref{sec5.1}.

It it noted that the energy-delay optimization for access control has never been studied in an online manner. This is due to that, before synchronization, the access delay or the energy consumption of an IoT device is hardly known from the perspective of a BS. To solve it, the BS can collect the reports from those access success IoT devices. But, these reports are generally outdated, due to that they can only be gathered after the delay occurs (i.e., the access success  IoT devices may already suffered a long delay or large energy consumption). As the dynamics of the system is Markovian over the frames, using these outdated information to maximize the future performance can be extremely intractable. Thus, the optimization task given in \eqref{eq1} becomes a generally intractable POMDP problem. The partial observation refers to the fact that the BS can only know the transmission receptions, but unable to know all other information related to the communication environment, including, but not limited to, the random collision occurrence as well as traffic statistics. Approximate solutions will be discussed in Sections \ref{sec4} and \ref{sec5}.

\vspace*{-0.2cm}
\subsection{System Model}\label{sec3.2}

In our system model, we consider all frames have the same length, and each frame contains $F$ available preambles. At the beginning of each frame, every IoT device can be activated according to a packets inter-arrival process. If activated, the IoT device receives the broadcast information from the BS, including the factors $\{\beta^t_\text{ACB}, \beta^t_\text{BO}, \beta^t_\text{TD}, \beta^t_\text{TB}\}$. According to the received information, the IoT device will determine how it executes RACH in the following frames. In the following, we describe the three main parts of system model: inter-arrival traffics, RACH procedure, and RACH schemes.

\subsubsection{Inter-Arrival Traffics}\label{sec3.2.1}
Considering a bursty traffic scenario, where massive IoT devices are recovered due to an emergency event, e.g., earthquake alarm and fire alarms, and try to establish synchronization with the BS. Every IoT device would be activated at any time $\tau$, according to a time limited Beta probability density function $p(\tau)$ given as \cite[Section 6.1.1]{3GPP2011Study}
\begin{align}\label{eq3}
p(\tau) = \frac{{{\tau^{{\alpha}-1}}{{(T - \tau)}^{{\beta}-1}}}}{{{{T}^{{\alpha} + {\beta} - 2}}\text{Beta}({\alpha} , {\beta} )}} ,
\end{align}
where $T$ is the total time of the bursty traffic, and $\text{Beta}({\alpha}, {\beta})$ is the Beta function with the constant parameters ${\alpha}$ and ${\beta}$ \cite{gupta2004handbook}. Considering that the newly activated devices at frame $t$ only come from those who received an packet within the interval between with the last RACH period ($\tau^{t-1}, \tau^t$), the packet inter-arrival rate measured in each RACH period at each IoT device is hence described by
\vspace*{-0.1cm}
\begin{align}\label{eq4}
\mu^t = \int_{{{\rm{\tau }}_{t-1}}}^{{\tau_{t}}} {p(\tau)} d\tau .
\end{align}

\subsubsection{Random Access Schemes}\label{sec3.2.3}

Once activated at frame $t$, each IoT device synchronizes to the broadcast timing and receives the broadcast from the BS. The received system information are used to determine their RACH frame according to the mechanism of the RACH schemes, which includes the information of the RACH access control factors $\{\beta^t_\text{ACB}, \beta^t_\text{TD}, \beta^t_\text{TB}, \beta^t_\text{BO}\}$. These factors are updated at each frame in order to adaptively alleviate the collisions during RACH. Each IoT device should schedule its RACH transmission intervals according to each RACH scheme based on these factors. In the following, RACH schemes will be detailed according to their execution order.
\begin{itemize}
    \item \textbf{ACB scheme}: $\beta^t_\text{ACB}$ is the ACB factor, which is defined as the probability of an IoT device to execute RACH in the current frame $t$. At the beginning of the current frame, each activated IoT device randomly generate a number $q$ between 0 and 1, and attempts to RACH only when $q ≤ \beta^t_\text{ACB}$, otherwise, the IoT device wait in this frame, and repeats the ACB check in the next frame.

\vspace*{-0.0cm}
\begin{figure}[htbp!]
    \begin{center}
        \includegraphics[width=0.8\textwidth]{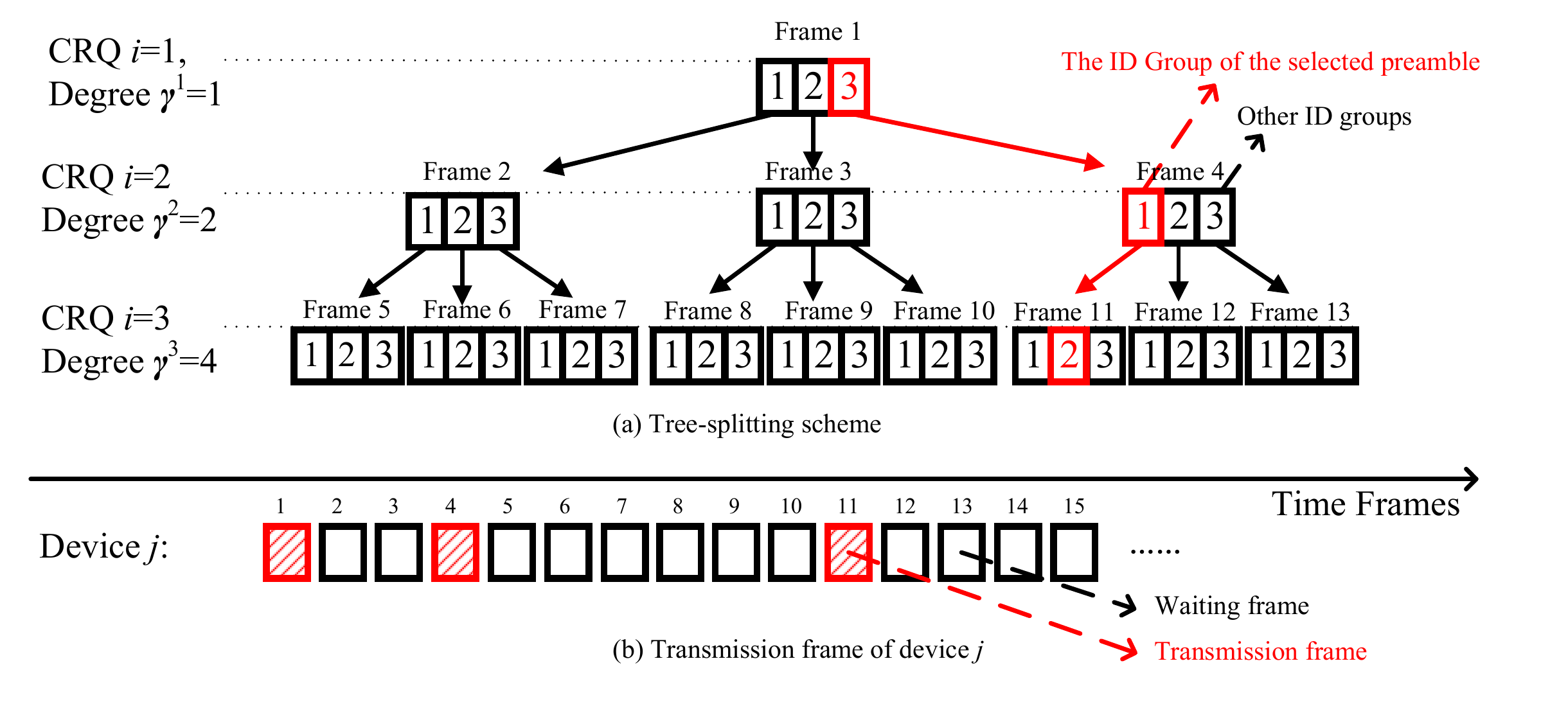}
         \caption{An example of the $3$-ary ($\beta^t_\text{Tdegree}=3$) DQ scheme with depth $\beta^t_\text{TDepth}=3$.}
         \label{figs2_0}
    \end{center}
\end{figure}

    \item \textbf{DQ scheme}: If the IoT device passed the ACB check, it would execute RACH immediately. If the current RACH fails, it will execute re-transmission according to a $\beta^t_\text{Tdegree}$-ary splitting-tree algorithm with finite $\beta^t_\text{TDepth}$ attempts. The degree $\beta^t_\text{Tdegree}$ refers to the number of branches that emanates from a tree node, while the depth $\beta^t_\text{TDepth}$ refers to the number of branches between a tree node and the root node. Thoroughly, each branch of the tree contains the same number of the preambles, where the total preambles are sequentially indexed $\{ 1, 2, \cdots, F \}$, and equally divided into $\beta^t_\text{TB}$ branches. For instance, assuming a $2$-ary tree with $F=54$ preambles, the first preamble group contains preambles with indexes from $1$ to $28$, and the second preamble group contains the others. 
    The IoT devices uniformly distributed among these preamble groups, due to the preamble random selection process. Based on the preamble group IDentification (ID), the collided IoT devices will be allocated to retransmit preamble in a specific future frame. For IoT device $j$, its retransmission frame is obtained by calculating a logical Collision Resolution Queue (CRQ) $i$, which includes two parameters: 1) the total length of the CRQ $\gamma^i = (\beta^t_\text{Tdegree})^i$ (number of nodes in depth $i$), and 2) the position of this IoT device $\mu_j^i$ in this CRQ. The IoT device $j$ can only transmit preamble at frame $\mu_j^i$ during the CRQ $i$. Denoting the prior preamble group IDs of $j$th IoT device as $\{\alpha_j^1, \alpha_j^2, \cdots, \alpha_j^{i-1}\}$, the position is calculated as 
    \vspace*{-0.1cm}
    \begin{align}\label{eq7}
    \mu_j^i =  \alpha_j^{i-1} + \sum_{k=1}^{i-2} (\beta^t_\text{Tdegree})^{i-1-k} (\alpha_j^k - 1) .
    \end{align}
    Once the $i$th CRQ finished, each collided IoT device would calculate their retransmission frame in the next CRQ $i+1$ based on \eqref{eq7}, until reaching the maximum depth $\beta^t_\text{TDepth}$. For better understanding, an example of $3$-ary splitting-tree is given in Fig. \ref{figs2_0}, where an IoT device first selects the $3$rd preamble group in the $1$st CRQ ($\{\alpha_j^1 =3$) for transmission, and then selects the $1$st preamble group in the $2$nd CRQ ($\alpha_j^2= 1$) for transmission. If unsuccessful, it select the $7$th preamble group (i.e., Frame 22) in the $3$rd CRQ for retransmission according to the calculation in \eqref{eq7}.

    \item \textbf{BO scheme}: After $\beta^t_\text{TB}$ attempts, the unsuccessful IoT device would switch to a back-off mode. These IoT devices postpone their access attempts for a period, whose length is uniformly selected from $[2^{\beta_\text{BO}-1}, 2^{\beta_\text{BO}}]$ frames\footnote{3GPP also supports back-off period within $[0, 2^{\beta_\text{BO}}]$ as shown in \cite{3GPP2017MAC}, while its performance would be worse than the proposed one if the RL-based optimization was used.}. After the back-off, the IoT device would listen to broadcasting, and re-attempt RACH according to the newly received system information.
    
\end{itemize}

For the ACB scheme, the BS is required to broadcast system information to all active devices in each frame. It can lead to relatively higher energy consumption, due to that if an IoT device's access request was denied, the energy it consumed to receive system information in this frame would be wasted. Conversely, the other two schemes can save more receiving energy, due to the direct allocation of re-transmissions frames for each IoT device. However, the ACB scheme will accurately control the traffic volume in each frame, thus is more capable in achieving a lower average access delay, whereas the DQ and BO schemes may allocate re-transmissions into a future frame that is heavily overloaded, and potentially lead to more collisions and higher access delay. Through tackling the problem in \eqref{eq1}, the access delay and energy consumption can be balanced without much sacrifice in the throughput.

\subsubsection{RACH Procedure}\label{sec3.2.2}

Each IoT device determines their RACH frame according to the received system information and the mechanism of the RACH schemes. During their RACH frame, one should execute a four-step RACH procedure immediately. From the perspective of an IoT device, the RACH procedure includes: 1) the IoT device transmits a randomly selected preamble (Msg1) from a pool with $F$ available preambles; 2) the IoT device waits to receive a Random Access Response (RAR) message (Msg2) within an RAR window; 3) if Msg2 was successfully received, the IoT device would send the Radio Resouce Control (RRC) connection request (Msg3) to the associated BS; 4) finally, the IoT device receives RRC connection confirmed message (Msg4) from the BS. 

The RACH procedure may fail due to the collision, which occurs when two or more IoT devices selecting the same preamble in Step 1. This collision comes from the fact that the BS cannot decode the Msg3 of RACH, due to the overlapped transmissions from collided devices using the same channel at the same time \cite{LTE2013dahlman}. To focus our study on the RACH schemes, the contention resolution is not considered in this paper. Thus we assume the BS cannot decode any collided signals as in prior works \cite{3GPP2011Study,he2017traffic,duan2016d,wu2012fast}. Based on the RACH model in 3GPP \cite{3GPP2011Study}, we assume the four-step RACH procedure, either in success or in collision, can finish within one frame (similar to \cite{duan2016d,azari2019latency,he2017traffic,lin2016estimation,wu2012fast}). Thus, if a preamble is collided, the related IoT devices can immediately re-attempt RACH in the next frame. However, each IoT device is allowed to execute RACH for a maximum number of attempts $\gamma_{\text max}$ \cite{3GPP2011Study}, otherwise, the packets in this device will be dropped.

To investigate the performance of RACH, we consider three KPIs, including the average throughput $V_s$, the average access delay $D$, and the average energy consumption $E$. The average access delay $D$ is averaging the total access delay over the number of success devices, where the delay is defined as the number of frames of the IoT device consumed from a newly packet arrival to the RACH success or fail (i.e., exceeds $\gamma_{\text max}$ RACH attempts). The average energy consumption $E$ is averaging the total energy consumption over the number of success devices calculated based on the system-level access model described in \cite[Sec. 7]{3GPP2015Cellular}. In details, the energy consumption of the $j$th IoT device to successfully access to the network is
\vspace*{-0.2cm}
\begin{align}\label{eq5}
E^j & =  E^j_\text{sy} + E^j_\text{RACH},
\end{align}
where $E^j_\text{sy}$ and $E^j_\text{RACH}$ are the energy consumption of the $j$th IoT device in receiving broadcast signal and executing RACH, respectively. According to \cite{3GPP2015Cellular}, $E^j_\text{sy}$ and $E^j_\text{RACH}$ can be obtained by calculating the product between their related average power consumption and working time. Therefore, Eq. \eqref{eq5} can be converted to 
\vspace*{-0.2cm}
\begin{align}\label{eq6}
E^j & =  n^j_\text{sy} T_\text{sy} P_\text{sy} + n^j_\text{RACH} (T_\text{Msg1} P_\text{Msg1} + T_\text{Msg2} P_\text{Msg2} + T_\text{Msg3} P_\text{Msg3} + T_\text{Msg4} P_\text{Msg4}),
\end{align}
where $n^j_\text{sy}$ and $n^j_\text{RACH}$ are the numbers of frames that the IoT device $j$ needed to receive broadcast signal and to execute RACH, respectively. In \eqref{eq6}, $T_\text{sy}$, $T_\text{Msg1}$, $T_\text{Msg2}$, $T_\text{Msg3}$, and $T_\text{Msg4}$ are constants, which are the consumed time in receiving broadcast signal, and in executing each step of RACH, respectively. Likewise, $P_\text{sy}$, $P_\text{Msg1}$, $P_\text{Msg2}$, $P_\text{Msg3}$, and $P_\text{Msg4}$ are also constants representing the average power consumption per time unit for receiving broadcast signal, and for executing each step of RACH, respectively. The effects of physical radio channels have already been considered in these constant factors \cite{3GPP2015Cellular}.

\section{Throughput Optimization Using Reinforcement Learning}\label{sec4}

Due to the complex long-term hybrid target defined in Eq. \eqref{eq1}, most prior works simply optimizing the number of access success devices, without taking into account the energy consumption and the access delay. Only optimizing the success access, the problem in \eqref{eq1} becomes 
\begin{align}\label{eq4-1}
\mathop {\text{max}}\limits_{ \pi(A^t|O^t)}   \sum_{k=t}^{\infty}  \gamma^{k-t} {V_s}^k,
\end{align}
or the even simpler one optimizing only the current frame
\begin{align}\label{eq4-2}
\mathop {\text{max}}\limits_{ \pi(A^t|O^t)} {V_s}^t.
\end{align}
The former one \eqref{eq4-1} has been considered in RL-based ACB scheme optimization as \cite{Luis2018Reinforcement}, while the latter simplified one \eqref{eq4-2} has been considered in most RACH optimization using conventional heuristic methods as \cite{wu2012fast,duan2016d,he2017traffic}. In order to evaluate the capability of RL algorithms, in this section, we design several RL algorithms to solve problem \eqref{eq4-1} with not only ACB scheme, but also DQ, and BO schemes, to be compared with the existing conventional heuristic methods. In the following, we first introduce the preliminary, including the perfect control strategy of the ACB scheme and the state-of-the-arts MLE-based ACB scheme. We then provide four state-of-the-arts RL algorithms, which are PG Reinforce and AC based on on-policy principle, and DQN and DDPG based on off-policy principle.

\subsection{preliminary}\label{sec4.1}

In this subsection, we introduce the conventional heuristic methods to optimize ACB scheme. Generally, the process of optimization is divided into two sub-tasks, which are traffic prediction and ACB factor configuration. The traffic statistic $\hat{N}^t$ is estimated based on the last observation $U^{t-1}=\{V_s^t, V_c^t, V_i^t\}$ obtained using MoM \cite{schoute1983dynamic,duan2016d,jiang2019online}, MLE\cite{he2017traffic}, or drift analysis \cite{wu2012fast}. Then, according to the estimated backlog $\hat{N}^t$, the optimized ACB factor $\beta^t_\text{ACB}$ for the next frame $t$ can be calculated by using $\beta^t_\text{ACB}=\text{min}(1,\frac{F}{\hat{N}^t})$ (proof can be found in \cite[Sec. IV.A]{duan2016d}), where $F$ is the number of available preambles. Next, we first review the current state-of-the-art conventional traffic estimator of \cite{he2017traffic}, which is based on MLE, and then describe the ideal ACB control scheme with known traffic, namely, Genie-aided ACB.

\subsubsection{Maximum Likelihood Estimator (MLE)}

In \cite{he2017traffic}, the traffic prediction problem is cast as a Bayesian probability inference problem, which calculates the probability for each possible traffic statistics at every frame $t$ based on the last observation $U^{t-1}=\{V_s^t, V_c^t, V_i^t\}$. Due to the intractability of the given problem, an ideal assumption\footnote{Note that this assumption is necessary to enable tractability of traffic prediction in most existing conventional heuristic traffic prediction methods \cite{wu2012fast,he2017traffic,duan2016d}.} that the current traffic load at frame $t-1$ (occurred) equals to the traffic load at the frame $t$ (requires to be predicted) is made. However, even with this ideal assumption, the optimal Bayes estimator is still intractable \cite{he2017traffic}. To solve it, \cite{he2017traffic} presented the maximum likelihood of the Bayes estimator with respect to each traffic load value $n$ under each possible observation $u$, which is given as 
\begin{align}\label{eq4-3}
 \hat{N}_\text{ML}^{t}= \hat{N}_\text{ML}^{t-1} = \mathop {\text{arg max}}\limits_{ n\in\{0,1,...,N_{\text{max}}\}}     \quad  {\mathbb P} \{ U^{t-1} | N^{t-1} =  n  \},
\end{align}
where $N_\text{max}$ is an upper bound on the traffic load statistics to enable implementation. Note that each probability value in ${\mathbb P} \{ U^{t-1} | N^{t-1} =  n  \}$ produces the likelihood of a value $n$ with an observation $u$.

To solve problem \eqref{eq4-3}, it is assumed that, at each frame $t$, each activated IoT device sequentially and independently chooses their preamble one after another, rather than choosing simultaneously in practice. This assumption does not change the uniformly selection principle of the random access, while the sequential selection process formulates a Markov chain to facilitate the calculation of likelihood ${\mathbb P} \{ U^{t-1} | N^{t-1} =  n  \}$. Under this assumption, the vector of likelihoods ${\mathbb P} \{ O^{t} | N^{t} =  n  \}$ for every $n$ can be obtained by calculating the steady-state probability vector of the formulated Markov chain. In the run-time, the traffic statistics can be obtained by selecting the one with the maximal likelihood ${\mathbb P} \{ U^{t-1} | N^{t-1} =  n  \}$ under the specific observation $U^{t-1}$. The details on the numerical procedure of MLE traffic prediction can be found in \cite{he2017traffic}.

\subsubsection{Genie-Aided ACB}:
We consider an ideal upper bound that the actual number of RACH requesting IoT devices $N^t$ is available at the BS, namely, Genie-Aided ACB scheme. The BS optimizes the ACB factor $\beta^t_\text{ACB}$ according to the real backlog ${N}^t$.

\subsection{Reinforcement Learning}\label{sec4.2}

DRL is one of the most capable methods to optimally solve complex POMDP problems, due to the reliance on the deep neural networks as one of the most impressive non-linear approximation functions \cite{sutton2017reinforcement}. The related DRL algorithms have been widely used in the dynamic optimization for wireless communication systems, e.g., \cite{jiang2019cooperative,jiang2019deep}. However, despite that these algorithms are generally model-free, a direct application of these general DRL approaches can not facilitate the optimal solution in all sorts of dynamic optimization problems wireless systems. For instance, a discrete-action DRL algorithm (e.g., DQN  \cite{mnih2015human}) may not be the best option to optimize the ACB scheme, due to its requirements in non-discrete control of the ACB factor ($\beta^t_\text{ACB}$). Therefore, we compare the state-of-the-arts DRL approaches to evaluate their capability in optimizing the number of access success devices of each RACH scheme.

We now introduce the general framework of DRL-based approaches to tackle problem \eqref{eq1}. To optimize the number of success devices for a RACH scheme, we consider a DRL agent deployed at the BS, which learns to choose appropriate actions progressively by exploring the environment. One DRL agent is responsible for an output variable $A^t$ (action), which represents the selected network parameter at the frame $t$. For instance, considering the ACB scheme, the output action is the ACB factor $\beta^t_\text{ACB}$ for the frame $t$. The selection of variable $A^t$ depends on the value $Q(S^t,A^t)$ or directly the policy $\pi$ according to the observed state $S^t$. By using a delayed reward ${R}^t  = V_{\rm s}^{t+1} $, the DRL agent updates its policy $\pi$ of action $A^t$, in an online manner, to progressively find the optimal solution of the RACH scheme for every state $S^t$. 

Unlike the conventional methods presented in Sec. \ref{sec4.1}, which only considers the single last observation $O^{t-1}$, the state variable $S^t$ of one DRL agent at frame $t$ consists of information in previous $T_o$ frames $S^t=[O^{t-T_o},O^{t-T_o+1},...,O^{t-1}]$. Including this historical information is due to that they can be useful to capture time-correlated features of the traffic generation mechanism and the RACH schemes. For instance, assuming the history length $T_o$ is long enough, one BS may recognize the wake-up duration of the periodical activated IoT devices.

To recognize patterns in temporal data, we employ an RNN, specifically a GRU network, to approximate the value function $Q(S^t, A^t; \bm{\theta})$ or the policy $\pi(S^t|A^t; \bm{\theta})$ of each DRL algorithm, where $\bm{\theta}$ represents the weights matrix of the GRU RNN. A many-to-one \textit{stateless}\footnote{Different from the \textit{stateless} implementation, the \textit{stateful} RNN does not need to re-initialize the memory at each training step, while its training progress is more resource-hungry and less stable, due to that the batch training is hard to be implemented \cite{zaremba2014recurrent}.} implementation of GRU RNN is adopted, where the historical observations in $O^{t}$ is sequentially fed into the network, and only the RNN with the last input $O^{t-1}$ is connected to the output layer for the further RACH parameter generation at frame $t$ (to be detailed in follows). During the implementation, note that the memory of GRU RNN needs to be re-initialized at each frame, and the historical length $T_o$ should generally be chosen according to the expected memory for time-correlation recognition.

The input of each DRL agent is the variables in state $S^t$; the intermediate layers are the introduced GRU RNN; while the output layer depends on different DRL algorithms. According to the unique training principle of each DRL algorithm, a loss function $L(\bm{\theta}^t)$ can be calculated to update the value function approximator $Q(S^t,A^t;\bm{\theta})$ (value-based algorithm) or the policy approximator $\pi(A^t| S^t;\bm{\theta})$ (policy-based algorithm). As the GRU RNN is used as the intermediate layers, we adopt standard Stochastic Gradient Descent (SGD) implemented via BackPropagation Through Time (BPTT) \cite{werbos1990backpropagation} for updating as
\begin{align}\label{eq8}
\bm{\theta}^{t+1} =  \bm{\theta}^{t}  -   \alpha \nabla L(\bm{\theta}^t),
\end{align}
where $\alpha$ is the learning rate, and the loss function $L(\bm{\theta}^t)$ for each DRL algorithms will be detailed as follows. Here, we consider four DRL algorithms, including PG, AC, DQN, and DDPG. The former two are basic on-policy algorithms, while the latter two are state-of-the-arts off-policy algorithms. The learning principle of these four algorithms are described as follows.

\subsubsection{PG Reinforce}\label{sec4.2.1}
In this algorithm, the DRL agent learns a parameterized policy $\pi$ to select actions without consulting a value function. The output layer consists of a Softmax non-linearity with $|\cal A|$ number of probability factors, where $|\cal A|$ represents the size of action space. The probability ${\mathbb P} \{ a^t = A^t | S^t, \bm{\theta}_\text{PG}^t  \}$ of each possible action $a$ is selected under the state $S^t$, which is parameterized by the weights $\bm{\theta}_\text{PG}^t$ at frame $t$. Recall that $\bm{\theta}_\text{PG}^t$ consists of both the GRU RNN parameters and the weights of the softmax layer. To train the policy $\pi(A^t| S^t;\bm{\theta}_\text{PG})$ (following \eqref{eq6}), the loss function $L_\text{PG}(\bm{\theta}_\text{PG}^t)$ is given as
\begin{align}\label{eq9}
 \nabla L(\bm{\theta}_\text{PG}^t) =   {\mathbb E}_{S^i,A^i,G^{i}} \left[ \gamma^i G^i \nabla_{\bm{\theta}} \text{ln} \pi(A^i |S^i;\bm{\theta}_\text{PG}) \right],
\end{align} 
where the expectation is taken with respect to a minibatch for $i\in\{t-M_r,...,t+1\}$ with size $M_r$, $G^t = \sum_{k=0}^{\infty} \gamma^k R^{t+k+1}$ is the return at frame $t$, and $\gamma$ is the discount rate. The implementation of PG Reinforce algorithm for RACH optimization is shown in \textbf{Algorithm \ref{a1}}.

\begin{algorithm}
\footnotesize
\SetKwData{Left}{left}
\SetKwData{This}{this}
\SetKwData{Up}{up}
\SetKwFunction{Union}{Union}
\SetKwFunction{FindCompress}{FindCompress}
\SetKwInOut{Input}{input}
\SetKwInOut{Output}{output}
\caption{On-policy PG Reinforce/AC algorithms} \label{a1}
\Input{Action space $|A|$, Operation Iteration $I$.}
\BlankLine 
Algorithm hyperparameters: learning rate $\lambda \in (0,1]$, discount rate $\gamma \in [0,1)$ \;
Initialization of the parameterized policy $\pi(s| a;\bm{\theta})$ and its state-value function $v(s;\bm{w})$ if needed\;
\For{Iteration $\leftarrow 1$ \KwTo $I$}{
Initialization of $S^0$ by executing a random action; \\
\For{$t \leftarrow 0$ \KwTo $T-1$}{
Select action $A^t$ according to the current policy $\pi(a| S^t;\bm{\theta})$; \\
The BS broadcasts the selected action $A^t$ , and backlogged IoT devices execute RACH; \\
The BS observes $S^{t+1}$, and calculates the related $R^{t} = V_{s}^{t}$; \\
Restore the tuple $S^t, A^t, R^{t}$; \\
}
Obtain the trajectory of an episode $S^0, A^0, R^0, S^1, \cdots, S^{T-1}, A^{T-1}, R^{T}$; \\
Calculate return $G^t$ for every frame $t$; \\
\begin{minipage}[t]{0.43\linewidth}%
\textbf{PG}: Calculate the loss $\nabla L_\text{PG}(\bm{\theta}_\text{PG}^t)$ using Eq. \eqref{eq9};
    \end{minipage}
\begin{minipage}[t]{0.43\linewidth}%
\textbf{AC}: Calculate the critic loss $\nabla L_\text{AC}(\bm{w}_\text{AC}^t)$ and actor loss $\nabla L_\text{AC}(\bm{\theta}_\text{AC}^t)$ using Eq. \eqref{eq10} and Eq. \eqref{eq11}, respectively; 
    \end{minipage}
 \\
Update the policy parameters $\bm{\theta}$ and state-value function $\bm{w}$ (AC algorithm).
}
\end{algorithm}

\subsubsection{AC}\label{sec4.2.2}
In this algorithm, a parameterized policy $\pi(A^t| S^t;\bm{\theta}_\text{AC})$ is learned to select actions called \textit{actor}, and a state-value function $v(S^t;\bm{w}_\text{AC})$ is learned to evaluate the action called \textit{critic}. The \textit{actor} follows the setting of PG Reinforce, where the output layer includes $|A|$ number of Softmax units to generate policy probabilities. The \textit{critic} is a state-value function with the output of one linear unit. The loss function for training the \textit{critic} $L_\text{AC}(\bm{w}_\text{AC}^t)$ is given as
\begin{align}\label{eq10}
 \nabla L(\bm{w}_\text{AC}^t) =  {\mathbb E}_{S^i,G^{i}} \left[  \gamma^t (G^i - v(S^i;\bm{w}_\text{AC})) \nabla_{\bm{w}_\text{AC}} v(S^i;\bm{w}_\text{AC}) \right],
\end{align} 
and the loss function for training the \textit{actor} $L_\text{AC}(\bm{\theta}^t)$ is given as
\begin{align}\label{eq11}
 \nabla L(\bm{\theta}_\text{AC}^t) = {\mathbb E}_{S^i, A^i,G^{i}} \left[ \gamma^i (G^i - v(S^i;\bm{w}_\text{AC}))  \nabla_{\bm{\theta}_\text{AC}} \text{ln} \pi(A^i |S^i;\bm{\theta}_\text{AC}) \right].
\end{align}
The implementation of AC algorithm for RACH optimization is shown in \textbf{Algorithm \ref{a1}}.

\subsubsection{DQN}\label{sec4.2.3}

In this algorithm, the DQN agent learns a state-action value function approximator $Q(S^t, A^t;\bm{\theta}_\text{DQN})$ to select the action, where the output layer is composed of linear units. The weights matrix $\bm{\theta}_\text{DQN}$ is updated in a fully online manner, which occurs along each frame, so as to avoid the complexities of eligibility traces. Applying a double DQN training principle \cite{lillicrap2015continuous}, the loss function $\nabla L(\theta_\text{DQN}^t)$ used to train the value function approximator $Q(S^t, A^t;\bm{\theta}_\text{DQN})$ is given as
\vspace*{-0.15cm}
\begin{align}\label{eq12}
 \nabla L(\theta_\text{DQN}^t) =  &  {\mathbb E}_{S^i,A^i,R^{i+1},S^{i+1}} \left[\left( R^{i+1} + \gamma  \mathop{\text{max}}\limits_{a\in{\cal A}} Q(S^{i+1}, a ; \bar{\theta}_\text{DQN}^t) -  Q(S^i, A^i;   \theta_\text{DQN}^t) \right) \nabla_{\theta_\text{DQN}} Q(S^i, A^i; \theta^t_\text{DQN})\right],
\end{align} 
where $\bar{\theta}_\text{DQN}^t$ is a so-called target value function. Note that $\bar{\theta}_\text{DQN}^t$ is initialized using the same network structure as the primary DQN $\theta_\text{DQN}$, which is only used to estimate the future value of the Q-function. The parameter of the target network $\bar{\theta}_\text{DQN}^t$ is partially copied from the primary value $\theta_\text{DQN}$ at each frame (implementation details are shown in \textbf{Algorithm \ref{a2}}). The hybrid use of the primary DQN and its target network forms the so-called double DQN. Note that, different from the PG and the AC algorithms given in Sec. \ref{sec4.2.1} and Sec. \ref{sec4.2.2}, the expectation here is taken with respect to a random minibatch. This minibatch is uniformly picked in random from a finite replay memory with size $M$. The implementation of DQN algorithm for RACH optimization will be detailed in \textbf{Algorithm \ref{a2}}.

\begin{algorithm}
\footnotesize
\SetKwData{Left}{left}
\SetKwData{This}{this}
\SetKwData{Up}{up}
\SetKwFunction{Union}{Union}
\SetKwFunction{FindCompress}{FindCompress}
\SetKwInOut{Input}{input}
\SetKwInOut{Output}{output}
\caption{Off-policy DQN/DDPG algorithms} \label{a2}
\Input{Action space $\cal A$, Operation Iteration $I$.}
\BlankLine 
Algorithm hyperparameters: learning rate $\lambda \in (0,1]$, discount rate $\gamma \in [0,1)$ \;
Initialization of the action-state value function $Q(s, a;\bm{\theta}_\text{DQN})$  for DQN, or the parameterized actor $\pi(s;\bm{\theta}_\text{DDPG})$ and critic $v(s, a;\bm{w}_\text{DDPG})$ for DDPG\;
\For{Iteration $\leftarrow 1$ \KwTo $I$}{
Initialization of $S^0$ by executing a random action; \\
\For{$t \leftarrow 0$ \KwTo $T-1$}{
\begin{minipage}[t]{0.43\linewidth}%
\textbf{DQN}: \lIf{$p^t_{\epsilon}<\epsilon$}{select a random action $A^t$ from $\cal A$}   
\hspace{+0.7cm} \lElse{Select $A^t= \mathop {\text{argmax}}\limits_{{a\in {\cal A}}} Q(S^t, a; \bm{\theta_\text{DQN}})$} 
    \end{minipage}
    \begin{minipage}[t]{0.43\linewidth}%
\textbf{DDPG}: Select $A^t= \pi(S^t;\bm{\theta}_\text{DDPG}) + {\cal N}^t$; 
    \end{minipage} \\
BS broadcasts action $A^t$, and backlogged IoT devices execute RACH;\\
BS observes $S^{t+1}$ and calculates $R^{t} = V_{s}^{t+1}$; \\
Storing transition $(S^{t}, A^t, R^{t}, S^{t+1})$ in replay memory $M$;\\
Sampling random minibatch of transitions $(S^{j}, A^j, R^{j}, S^{j+1})$ from replay memory $M$; \\
\begin{minipage}[t]{0.43\linewidth}%
\textbf{DQN}: Calculate the loss of Q-function $\nabla L(\theta_\text{DQN}^t)$ \\
 $\vphantom{9}$ $\vphantom{9}$ $\vphantom{9}$ $\vphantom{9}$ $\vphantom{9}$     $\vphantom{9}$ $\vphantom{9}$ $\vphantom{9}$ using Eq. \eqref{eq12}; 
    \end{minipage}
    \begin{minipage}[t]{0.43\linewidth}%
\textbf{DDPG}: Calculate the loss of critic $\nabla L_\text{DDPG}(\bm{w}_\text{DDPG}^t)$ and actor $\nabla L_\text{DDPG}(\bm{\theta}_\text{DDPG}^t)$ using Eq. \eqref{eq13} and Eq. \eqref{eq14}, respectively;
    \end{minipage} \\
Perform a gradient descent for each primary network; \\
Update the target networks using: $\bm{\bar{w}}^t \leftarrow \sigma\bm{w}^t + (1-\sigma)\bm{\bar{w}}^t $, and $\bm{\bar{\theta}}^t \leftarrow \sigma\bm{\theta}^t + (1-\sigma)\bm{\bar{\theta}}^t $.
}
}
\end{algorithm}

\subsubsection{DDPG}\label{sec4.2.4}
This algorithm adapts the ideas of DQN to the continuous action selection. Rather than discrete action control in the BO and the DQ schemes, the DDPG can be applied for continuous action control in ACB access scheme. Similar to AC algorithm, DDPG leverage an \textit{actor} $\pi(S^t;\bm{\theta}_\text{DDPG})$ to learn action selection, while the \textit{critic} uses a state-action value function $v(S^t, A^t;\bm{w}_\text{DDPG})$ to evaluate the action. The input of \textit{critic} includes both the current state $S^t$ as well as the current action $A^t$ (obtained by the \textit{actor} $\pi(S^t;\bm{\theta}_\text{DDPG})$), and the output is a linear unit. In terms of the action selection, the \textit{actor} deterministically maps a state to a specific action according to the current policy $\pi(S^t;\bm{\theta}_\text{DDPG})$. To do so, the output of \textit{actor} uses a Sigmoid non-linearity unit, which generates continuous numbers within $(0,1)$. The Sigmoid output perfectly matches the range of the ACB factor $(0,1]$, thus it can be directly used as the ACB factor at each frame $t$. During the training, the \textit{critic} $L_\text{DDPG}(\bm{w}_\text{DDPG}^t)$ is updated by minimizing the loss using
\begin{align}\label{eq13}
 \nabla L(\bm{w}_\text{DDPG}^t) =  {\mathbb E}_{S^i,A^{i}, R^{i+1}, S^{i+1}} \Big[\big(   & R^{i+1} + \gamma  v(S^{i+1}, \bar{\pi}(S^i;\bar{\bm{\theta}}_\text{DDPG}) ; \bar{\bm{w}}_\text{DDPG}^t) -  
 \nonumber \\
 & v(S^i, A^i;  \bm{w}_\text{DDPG}^t) \big) \nabla_{\bm{w}_\text{DDPG}} v(S^i, A^i; \bm{w}^t_\text{DDPG})\Big],
\end{align} 
where $\bar{\bm{w}}_\text{DDPG}$ and $\bar{\bm{\theta}}_\text{DDPG}$ are the weights of the target critic and actor, respectively. Similar to DQN, these two target networks are updated by partially copying from the primary two. To train the \textit{actor} $L_\text{DDPG}(\bm{\theta}^t)$, the loss function is given as
\begin{align}\label{eq14}
 \nabla L(\bm{\theta}_\text{DDPG}^t) = {\mathbb E}_{S^i, A^i,R^{i+1}, S^{i+1}} \left[ \nabla_{a} v(S^i, a; \bm{w}^t_\text{DDPG})|_{a=\pi(S^i) } \nabla_{\bm{\theta}_\text{DDPG}} \pi(S^i;\bm{\theta}_\text{DDPG}) \right].
\end{align}
In \eqref{eq13} and \eqref{eq14}, the expectation is also taken using a random minibatch. The implementation of DDPG algorithm for RACH optimization is shown in \textbf{Algorithm \ref{a2}}.

\subsection{Numerical Results and Evaluation}\label{sec4.3}

\begin{figure}[htbp!]
    \begin{center}
    \subfigure[\scriptsize $T = 20$ frames]{
        \includegraphics[width=0.48\textwidth]{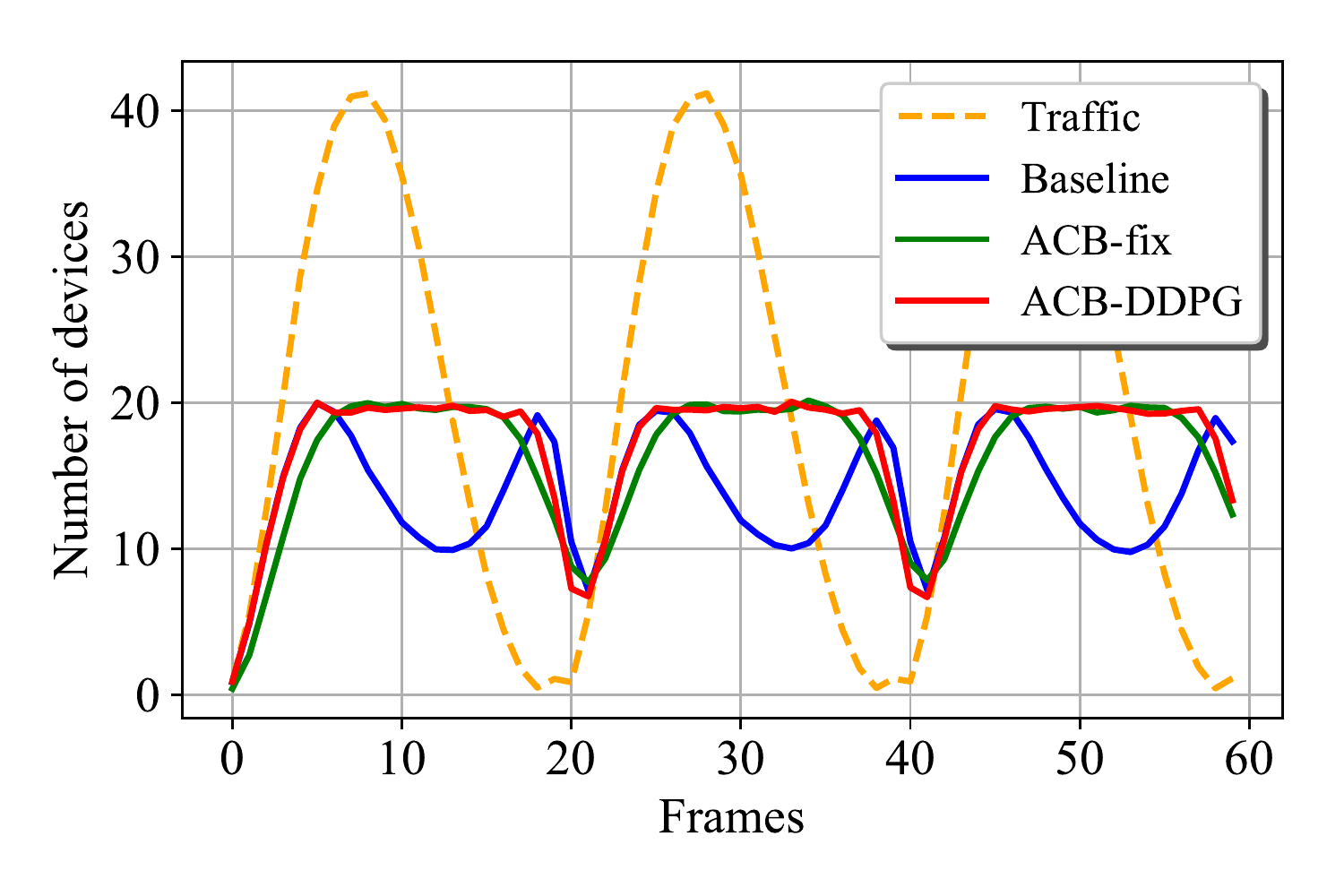}}
        \subfigure[\scriptsize $T = 10$ frames]{ \includegraphics[width=0.48\textwidth]{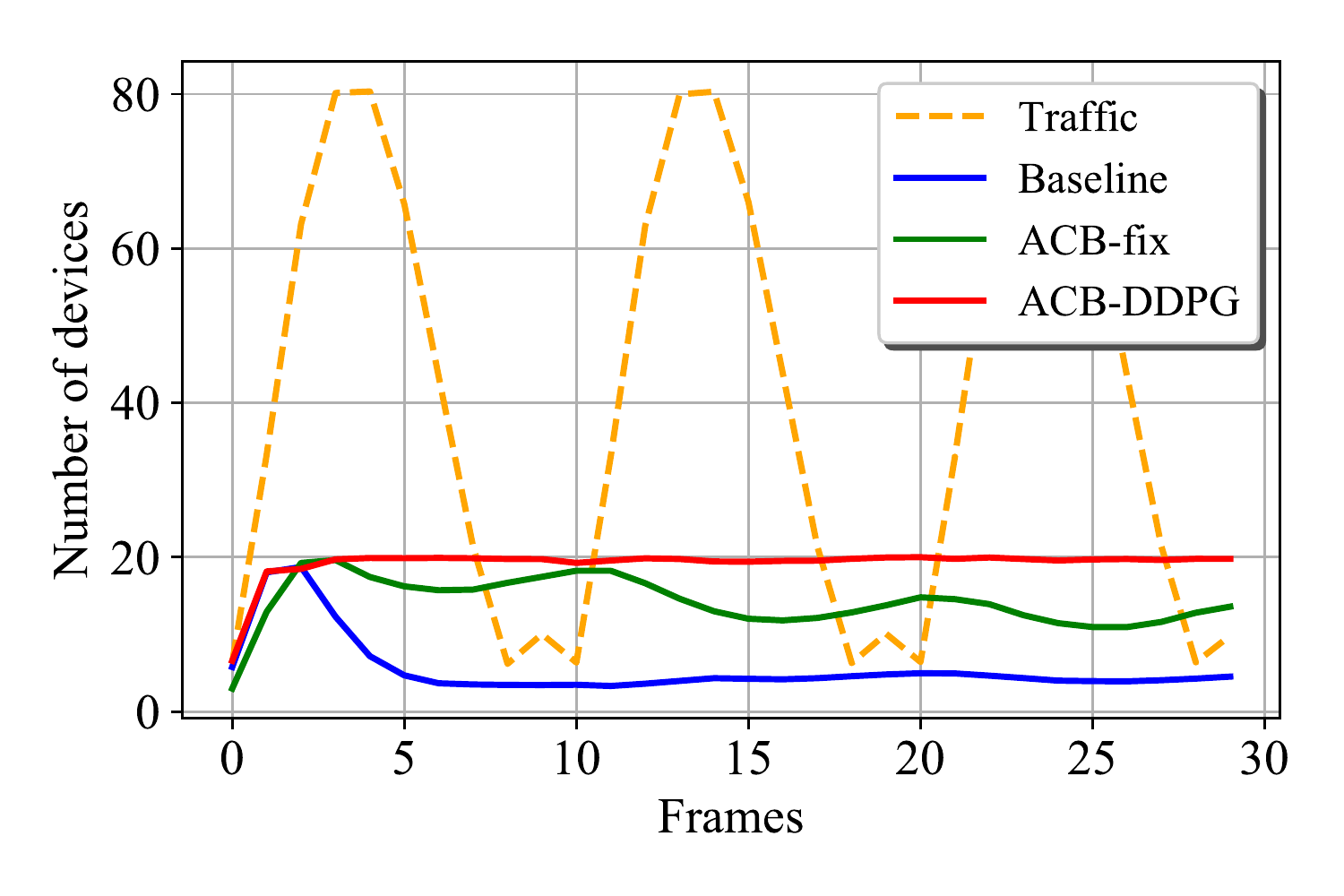}}
         \caption{The average number of successfully accessed IoT devices $V_\text{s}$ per frame over 1000 epochs, where (a) uses the time limited Beta traffic profile with period $T = 20$, and (b) uses the similar traffic profile with period $T = 10$. The dashed line represents the average number of IoT devices with new arrival packets per frame.}
         \vspace*{-1.4cm}
         \label{figr1}
    \end{center}
\end{figure}

In this subsection, we evaluate the number of access success devices of the ACB, BO, and DQ schemes using our proposed four DRL algorithms above via numerical experiments. We adopt the standard network parameters following 3GPP technical report for MTC systems \cite{3GPP2011Study,3GPP2015Cellular}, where the number of preambles $F=54$, retransmission constraint $\gamma_\text{max}=10$, and each frame contains $640$ milliseconds (ms). Unless otherwise stated, we mostly assume the presence of devices $N=400$ generating a packet at random according to the time limited Beta profile with parameters $({\alpha} , {\beta} )=(3,4)$ with period $T = 20$. An example of the resulting average number of activated devices of this Beta profile is shown as orange dashed line in Fig. \ref{figr1}. For different DRL algorithms, we use the same hyperparameters for the training as well as the testing for fair comparison. Each DRL agent is with three layers, where the first two layers are each with 128 GRU units, and the last layer is with 128 Rectifier Linear Units (ReLUs). The other hyperparameters can be found in Table \ref{table1}. The action of ACB $\beta^t_\text{ACB}$ is a factor selected from $(0, 1]$, where the discrete model selects factor with the minimum pace of $0.05$, and the continuous one can select any value without any limitation. The BO and DQ schemes can only be controlled in a discrete manner, where the action of BO $\beta^t_\text{BO}$ is an integer selected from $[0, 8]$, and the DQ scheme can use any tree smaller than $\{\beta^t_\text{Tdepth}, \beta^t_\text{Tdegree}\} = \{3,6\}$.

                        

\begin{table}[htbp!]
\setcounter{table}{0}
	\centering
	\caption{RL Hyperparameters \vspace*{-0.0cm}}
	{\renewcommand{\arraystretch}{0.6}
		\begin{tabular}{|*{1}{p{5cm}}|*{1}{p{2cm}} |*{1}{p{5cm}}|*{1}{p{2cm}} |}
			\hline
			\rowcolor{Gray}
		   \bf{Hyperparameters}   &    \bf{Value}$\vphantom{\Big(}$   & \bf{Hyperparameters}   &    \bf{Value}$\vphantom{\Big(}$  \\  \hline
		   $\vphantom{\big(}$Memory $T$ & 20 & 
		   $\vphantom{\big(}$Adam learning rate $\alpha$ & 0.0001 \\
           $\vphantom{\big(}$Minibatch size $\vphantom{\big(}$ & 32 & 
           $\vphantom{\big(}$Historical samples size $M$ & 10000 \\
           $\vphantom{\big(}$Minimal exploration rate $\epsilon$  & 0.01 & 
           $\vphantom{\big(}$ACB scheme discount rate $\gamma_\text{ACB}$  & 0.1 \\
           $\vphantom{\big(}$BO and DQ schemes discount rate $\gamma$  & 0.9 & 
           $\vphantom{\big(}$Target network update rate $\sigma$  & 0.2 \\
                        \hline
                        
		\end{tabular}
	}
	\vspace*{-0.1cm}
	\label{table1}
\end{table}

\vspace*{-0.0cm}

We first illustrate the operation of the proposed traffic and ACB scheme under the time limited Beta traffic profile with period $T = 20$ in Fig. \ref{figr1}(a) and period $T = 10$ in Fig. \ref{figr1}(b). Fig. \ref{figr1} plots the average number of IoT devices under new arrival packets in each frame, and the average number of the access success devices $V_\text{s}$ per frame for the baseline scheme, the fixed-factor ACB scheme, as well as the DDPG-based ACB scheme, respectively. The baseline refers to transmissions without any access control, the fixed ACB scheme uses the fixed factor $\beta^t_\text{ACB}=0.5$, and the DDPG-based ACB scheme relies on the DDPG algorithm to adaptively assign the ACB factor in each frame as given in Sec. \ref{sec4.2.2}. We first observe that, with the smaller traffic arrival period, the new arrival packets in Fig. \ref{figr1}(b) is much more intense than that in Fig. \ref{figr1}(a). In Fig. \ref{figr1}(a), it is observed that the access performance of the baseline scheme (i.e., without ACB control) reduces dramatically after the new arrival traffic hits the peak. This is due to that the number of access requests during this period has been accumulated to the maximum, which increases collisions. In contrary, the access performance of fixed-factor ACB schemes remain at around $20$ during that period, but is worse than that of the baseline scheme when the traffic is small. This is because the access baring mechanism reduces collisions during the peak traffic period, but also decreases the preamble utilization during the non-peak traffic period. Different from Fig. \ref{figr1}(a), the access performance in Fig. \ref{figr1}(b) always follow ACB-DDPG$>$ACB-fix$>$Baseline, due to its much heavier incoming traffic. It can be seen that, in both figures, the DDPG-based ACB scheme performs the best at both peak and non-peak periods, which is due to the capability of ACB-DDPG in assigning ACB factor in an online manner to adaptively control the number of access devices in an acceptable level.

\begin{figure}[b!]
    \begin{center}
    \begin{minipage}[t]{0.47\textwidth}
        \includegraphics[width=1\textwidth]{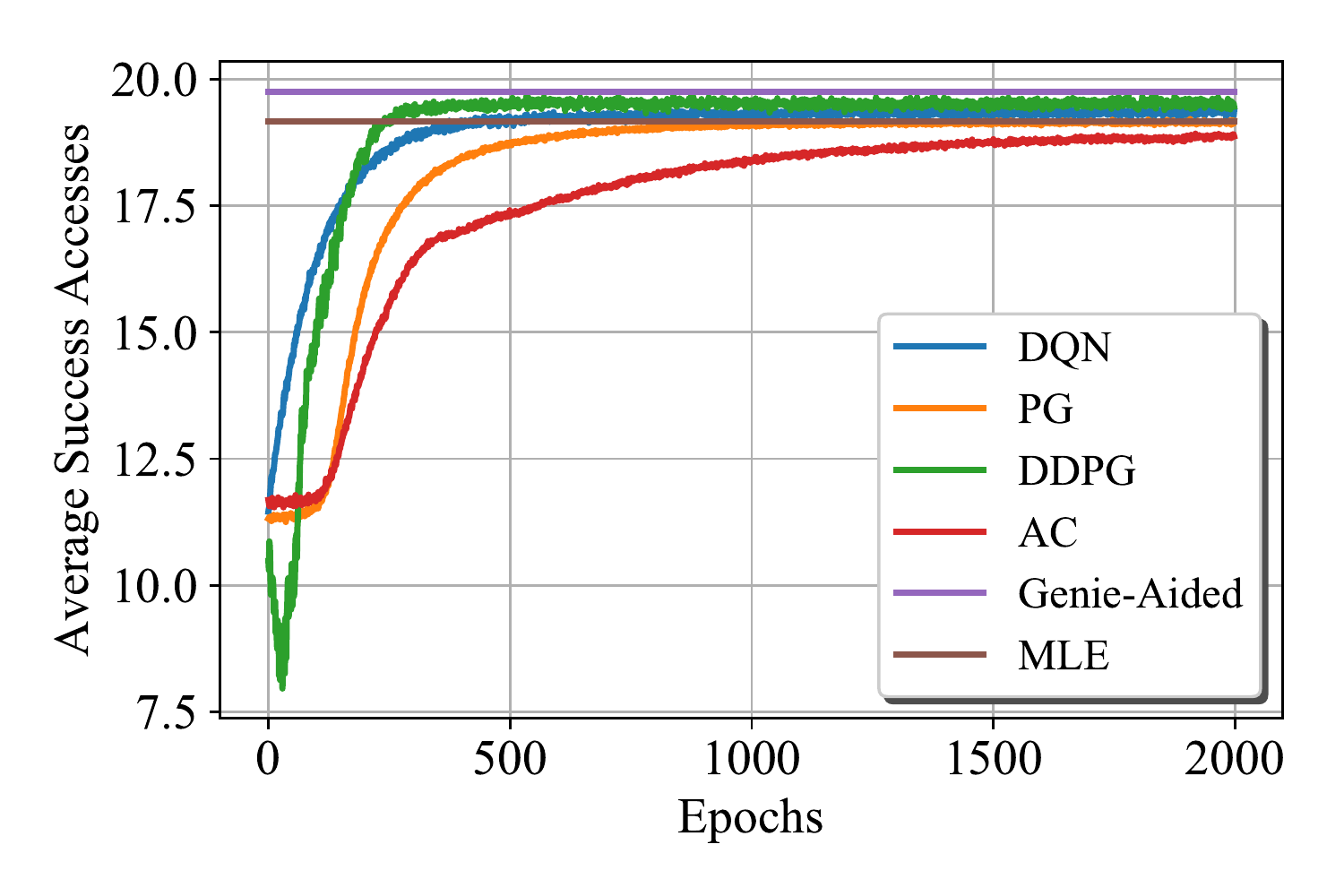}
        \vspace*{-1.2cm}
         \caption{
          Average number of successfully accessed IoT devices per frame as a function of the frames in the online training phase.
          }
          \vspace*{-0.8cm}
         \label{figr2}
    \end{minipage}
      \hspace*{+0.5cm}
        \begin{minipage}[t]{0.47\textwidth}
    \centering
        \includegraphics[width=1\textwidth]{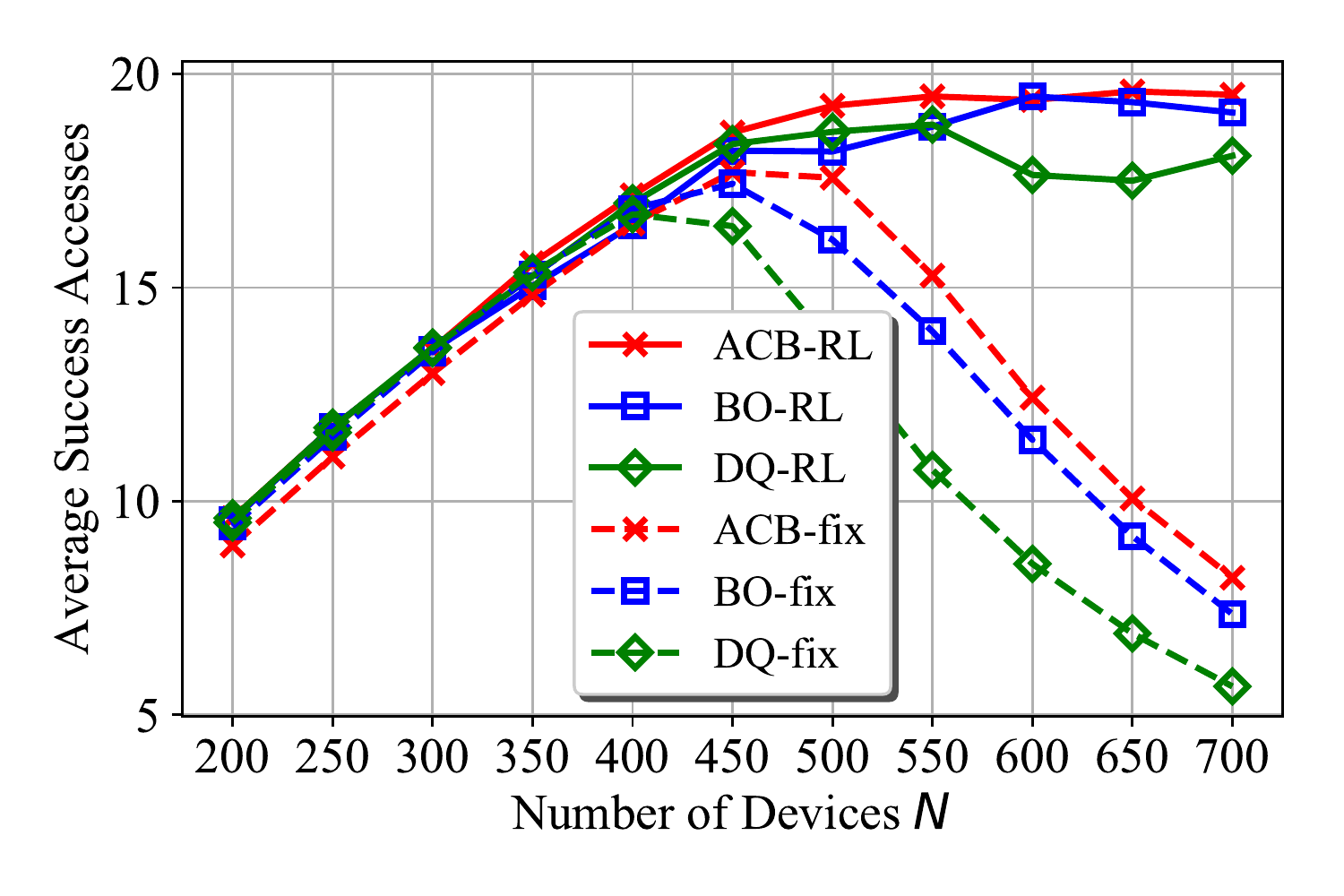}
        \vspace*{-1.2cm}
         \caption{The average number of successfully accessed IoT devices $V_\text{s}$ during 1000 testing episodes.}
                \label{figr3}
        \end{minipage}
    \end{center}
    \vspace*{-0.8cm}
\end{figure}

Fig. \ref{figr2} compares the evolution of the average number of the access success devices per episodes for each RL algorithm in the training phase under the ACB scheme, where each result is averaged over 100 training trails. The training curves are also compared with lower and upper bounds using the MLE-based ACB scheme and the ideal genie-aided ACB scheme, respectively. It can be seen that, after convergence, the proposed DDPG and DQN methods are quite close to the ideal upper bound, and slightly outperform the conventional MLE-based ACB scheme. This improvement (around $2.5\%$) is relatively minor, due to that the optimization of the ACB scheme is mathematically tractable using MLE. In fact, the ACB factor would be optimally selected, if the actual backlog was given (according to Sec. \ref{sec3.2}). Despite that the improvement of ACB scheme is relatively minor, we argue that the use of RL algorithms in RACH optimization is still worthwhile. The reason is that the optimization of the other RACH schemes are mathematically intractable, such as the presented BO and DQ schemes.

\vspace*{-0.0cm}

In Fig. \ref{figr2}, it is seen that the off-policy algorithms (DDPG and DQN) outperform the other on-policy algorithms (PG and AC) in terms of both the adaptation speed and the performance after convergence. This is mainly due to the experience replay mechanism, which efficiently utilizes the training samples, and also due to the randomly sampling, which smooths the training distribution over the previous behaviors. We further observes that, despite the similar learning principle and neural network architecture used, the DDPG slightly outperforms the DQN. This is due to the continuous control mechanism of DDPG, which provides the infinite action space of the ACB configuration, whereas the DQN is only limited to the configured $20$ action space. Motivated by this, we use DDPG for ACB configuration and DQN for any other discrete factors configurations, including BO and DQ in the rest of figures. Furthermore, in practice, one important reason to use off-policy RL algorithms in RACH optimization is that the training can be implemented in a cloud or an edge by gathering samples from multiple BSs, which would greatly save computational resource. Note that this implementation is based on the assumption that the samples of different BSs are identically and independently distributed.


Fig. \ref{figr3} plots the average number of the access success devices per frame per episode using the fixed-factor RACH schemes and RL-based RACH schemes versus various numbers of devices $N$. The ``ACB-fix", ``BO-fix", and ``DQ-fix" are the ACB, BO, and DQ schemes based on the fixed factors of $\beta^t_\text{ACB}=0.5$, $\beta^t_\text{BO}=2$ (back-off from $[2, 4]$ frames), and $\{\beta^t_\text{TD},\beta^t_\text{TB}\}=\{2,2\}$, respectively. It is observed that all schemes achieve similar performance when the number of devices $N$ is smaller than $400$, but the RL-based schemes substantially outperform the conventional fixed-factor (non-dynamic) access control schemes in heavy traffic region, where the number of devices $N$ is bigger than $400$. This showcases the capability of RL algorithms to better optimize each RACH scheme, as they can well manage access load in the presence of heavy traffic. It is also interesting to note that ``ACB-RL" generally outperforms the ``BO-RL" and ``DQ-RL" in any number of devices. This is because the ACB schemes can control the access request probability of each device in each frame, whereas the ``BO-RL" and ``DQ-RL" may allocate re-transmissions into a future frame that is heavily overloaded, which may lead to more collisions. 


\section{Hybrid Scheme Optimization and Decoupled Learning Strategy}\label{sec5}

In this section, we aims at solving problem \eqref{eq1} and jointly optimizes three RACH schemes defined by factors $A^t=\{\beta^t_\text{ACB}, \beta^t_\text{BO}, \beta^t_\text{Tdepth}, \beta^t_\text{Tdegree}\}$. This hybrid scheme integrates the features of the ACB, the BO, and the DQ schemes, where they may be jointly executed according to control of the DRL agents. Different from the single scheme scenario in Sec. \ref{sec4}, this joint execution increases adjustable system factors, which results in an exponential increment of the action space. As evaluated in our previous work \cite{jiang2019deep}, optimizing a system with such numerous action space by a single RL agent is not feasible, due to the convergence difficulty. To solve problem \eqref{eq1}, in this section, we propose a decoupled learning strategy to efficiently train multiple parallel RL agents that each handles one access control factor. In the following, we first introduce the basic multi-agent cooperative RL for multi-factors optimization. After that, we propose the decoupled learning strategy in detail.

\vspace*{-0.0cm}
\begin{figure}[htbp!]
    \begin{center}
    \subfigure[\scriptsize Conventional multi-agent cooperative DRL structure.]{
        \includegraphics[width=0.9\textwidth]{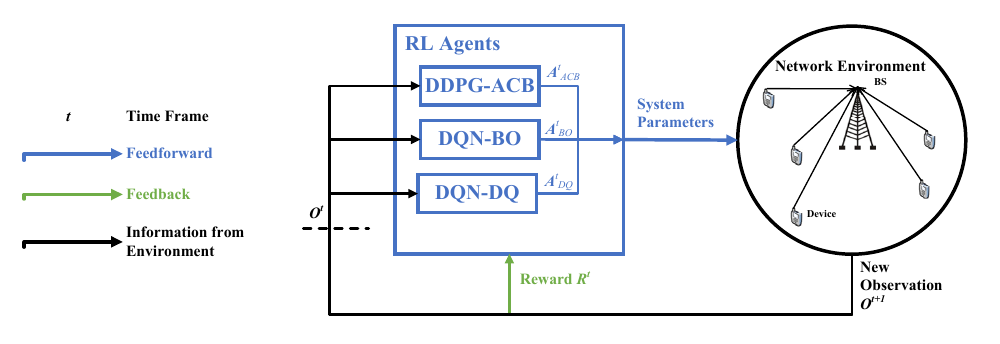}}
        \subfigure[\scriptsize Decoupled learning strategy for multi-agent cooperative DRL.]{ \includegraphics[width=0.9\textwidth]{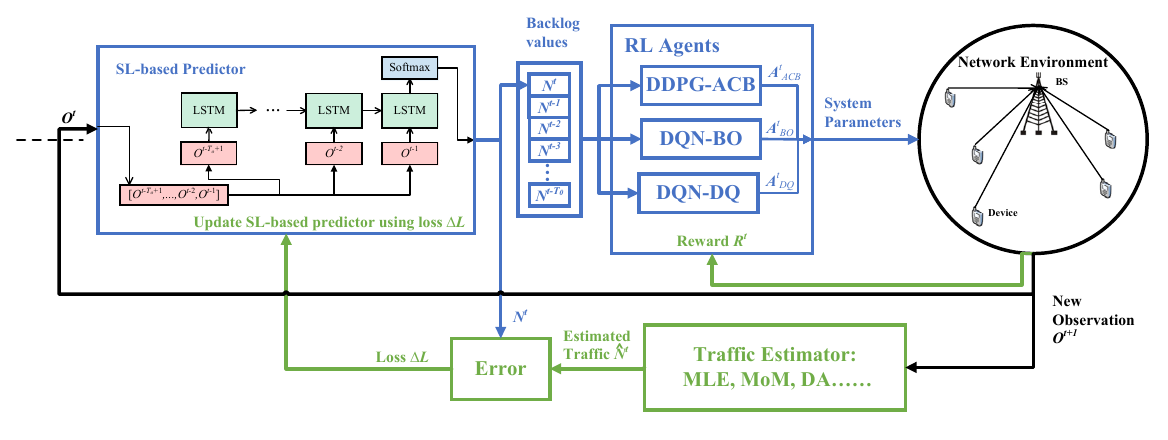}}
         \caption{Proposed learning structure of (a) conventional multi-agent cooperative DRL, and (b) decoupled multi-agent cooperative DRL.}
         \vspace*{-0.8cm}
         \label{fig5}
    \end{center}
\end{figure}

\subsection{Multi-Agent Cooperative Deep Reinforcement Learning}\label{sec5.1}

In this subsection, we introduce a basic multi-agent cooperative DRL method to tackle problem \eqref{eq1}. As evaluated, a direct application of DQN or DDPG given in Sec. \ref{sec4} is not feasible to solve this problem, due to the enormous size of the action $A^t$. To solve this problem, the action space $A^t=\{\beta^t_\text{ACB}, \beta^t_\text{BO}, \beta^t_\text{Tdepth}, \beta^t_\text{Tdegree}\}$ can be broken down into three separate sub-actions, each controls one scheme, including $A^t_\text{ACB}$, $A^t_\text{BO}$, and $A^t_\text{DQ}$. As illustrated in Fig. \ref{fig5} (a), we consider three independent RL agents to handle each of these actions, where the first, $A^t_\text{ACB}$, is handled by a DDPG agent, and the other two are handled by two DQN agents. Each agent parameterizes their own value function $Q(S^t,A^t)$  or policy $\pi$ by using a function $\theta$, which is represented by the multiple layers GRU RNN structure given in \ref{sec4.2}.

In each frame, the state variable $S^t=[O^{t-T_o},O^{t-T_o+1},...,O^{t-1}]$ is fed into each agent to generate sub-actions, where each observation $O$ not only includes the historical transmission receptions $U$, but also the historical action selection of every agent $A=\{A_\text{ACB}, A_\text{BO}, A_\text{DQ}\}$. The share of historical action selection among each agent aims to benefit their cooperation. By doing so, each agent can understand how the total reward is influenced by each sub-action, so as to predict the future action selection of each other. After each frame, DRL agents are trained in parallel, where their parameters $\bm{\theta}$ and/or $\bm{w}$ are updated using the approaches given in Eq. \eqref{eq9}-\eqref{eq14} of Sec. \ref{sec4.2}. Note that this cooperative training approach has been proposed in our previous works \cite{jiang2019cooperative,jiang2019deep}, and is akin to theproportional approach proposed in \cite{Sunehag2018}, which has been evaluated as a close replacement of the overall function by a factorization.

The training of each DRL agent shares a common reward signal, which guarantees that all of them aims at the same objective as given in Eq. \eqref{eq1}. The reward is obtained by using the weighted sum of the success accesses reward $R_{\rm s}^t$, the access delay reward ${R_{\rm d}^t}$, and the energy consumption reward ${R_{\rm e}^t}$ as given in Eq. \eqref{eq2}. The success accesses reward $R_{\rm s}^t$ is derived by directly normalizing the observed average success accesses $V_{\rm s}^t$. On contrary, the access delay reward ${R_{\rm d}^t}$ and the energy consumption reward ${R_{\rm e}^t}$ are inversely to the average access delay reward ${V_{\rm d}^t}$ of each succeeded device and the average energy consumption ${V_{\rm e}^t}$, which are derived by using a revised hyperbolic tangent activation function (a.k.a., tanh function) as
\vspace*{-0.1cm}
\begin{align}\label{eq17-2}
R^{t} = 1 - \frac{e^{\frac{V^t}{c}} - e^{-\frac{V^t}{c}}} {e^{\frac{V^t}{c}} + e^{-\frac{V^t}{c}}},
\end{align}
where $c$ is a constant factor. In \eqref{eq17-2}, the use of the revised tanh function targets to not only generating the sub rewards ${R_{\rm d}^t}$ and ${R_{\rm e}^t}$ that is inversely proportional to observations ${V_{\rm d}^t}$ and ${V_{\rm e}^t}$, but also limiting the sub rewards in the range of $(0, 1]$ to improve the convergence capability of neural networks. Note that the constant factor $c$ is used to flexibly determines the density of reward, where its selection impacts on both the convergence capability and the training efficiency of neural networks.

\subsection{Decoupled Learning Strategy for Hybrid Scheme Optimization}\label{sec5.2}

Applying conventional DRL methods to optimize massive access control faces two challenges: \textit{i)} the DRL agent is expected to be updated in an online manner, but the convergence is really slow due to the complexity of neural network as well as the tradeoff between exploration and exploitation; and \textit{ii)} the access performance of the converged DRL agents degrades by various hidden information, including, but not limited to, the traffic arrival pattern and random collision occurrence. Aiming at tackling these two challenges, in this subsection, we propose a decoupled learning strategy to train multiple parallel DRL agents as shown in Fig. \ref{fig5}(b). Different from conventional DRL, problem \eqref{eq1} was decoupled into two sub-tasks, including traffic prediction and parameter configuration. These two sub-tasks are successively solved, where we first predict the traffic by using an online supervised learning method \cite{jiang2019online}, and we then input the predicted traffic statistic into DRL agents to configure actions. Details will be discussed in following two parts.

\subsubsection{Online Supervised Learning for Traffic Prediction}

This learning-based predictor adopts a modern RNN model with parameters $\bm{\theta}_\text{RNN}$ to predict the traffic statistic at each frame. The use of RNN is due to its capability in capturing time correlation of traffic statistics over consecutive frames, which can help to learn the time-varying traffic trend for better prediction accuracy. As shown in Fig. \ref{fig5}(b), historical observations $O_{t-T_o}^{t-1}$ are sequentially inputted into the RNN predictor in a \textit{stateless} manner, and only the RNN with the last input $O^{t−1}$ is connected to the output layer. The output layer consists of a Softmax non-linearity with ($N_\text{max}$+1) values, which represent the predicted probability ${\mathbb P} \{ \hat{N}^{t} = n |O_{t-T_o}^{t-1}, \bm{\theta}_\text{RNN}\}$ of each possible value $n$ for the forthcoming backlog $N^{t}$. Recall that $N_\text{max}$ is the upper bound on the backlog statistics to ease implementation. In order to enable online updating, we also implement the label estimation method proposed in \cite{jiang2019online}. Briefly speaking, with respect to a produced backlog $\hat{N}^{t}$ at any frame $t$, the weights of the RNN are adapted with a delay of one frame at frame $t+1$ by using an estimated label $\tilde{N}^{t}$. This label $\tilde{N}^{t}$ can be obtained at the next frame $t+1$ by using the MLE estimator described in Sec. \ref{sec3.1}, or MoM estimator described in \cite{duan2016d,jiang2019online}. At each frame $t+1$, the parameters $\bm{\theta}_\text{RNN}$ of the RNN predictor are updated using SGD via BPTT as
\begin{align}\label{eq18}
\bm{\theta}_\text{RNN}^{t+1} =  \bm{\theta}_\text{RNN}^{t}  -   \alpha \nabla L(\bm{\theta}_\text{RNN}^t),
\end{align}
where $\nabla L(\bm{\theta}_\text{RNN}^t)$ is the gradient of the loss function $L(\bm{\theta}_\text{RNN}^t)$ to train the RNN predictor. This is obtained by averaging the cross-entropy loss as
\begin{align}\label{eq19}
 L^{t}({\bm{\theta}_\text{RNN}}) =    - \sum_{t'=t-T_b+1}^{t} \text{log}\left({\mathbb P} \{ \hat{N}^{t'} = \tilde{N}^{t'} | O_{t'-T_o}^{t'}, \bm{\theta}_\text{RNN}  \}\right),
\end{align} 
where the sum is taken with respect to randomly selected mini-batch with size $T_b$.

\begin{algorithm}
\footnotesize
\SetKwData{Left}{left}
\SetKwData{This}{this}
\SetKwData{Up}{up}
\SetKwFunction{Union}{Union}
\SetKwFunction{FindCompress}{FindCompress}
\SetKwInOut{Input}{input}
\SetKwInOut{Output}{output}
\caption{Decoupled learning strategy for multi-agent DRL training} \label{a3}
\Input{Action space $\cal A_\text{ACB}$, $\cal A_\text{BO}$, $\cal A_\text{DQ}$, and operation iteration $I$.}
\BlankLine 
Algorithm hyperparameters: learning rate $\lambda_\text{RNN} \in (0,1]$ for traffic predictor, and learning rate $\lambda_\text{DRL} \in (0,1]$, as well as discount rate $\gamma \in [0,1)$ for DRL agents;\\
Initialization of the RNN parameters $\bm{\theta}_\text{RNN}$ for traffic predictor, the action-state value function $Q(s, a;\bm{\theta}_\text{BO})$ for the BO scheme, the action-state value function $Q(s, a;\bm{\theta}_\text{DQ})$ for the DQ scheme, and the parameterized actor $\pi(s;\bm{\theta}_\text{DDPG})$ as well as critic $v(s, a;\bm{w}_\text{ACB})$ for the ACB scheme;\\
\For{Iteration $\leftarrow 1$ \KwTo $I$}{
Initialization of $O^0$ by executing a random action; \\
\For{$t \leftarrow 0$ \KwTo $T-1$}{
Predict backlog value $\hat{N}^{t}$ using $\bm{\theta}_\text{RNN}^{t}$ according to the historical observations $[O^{t-T_o+1},\cdots,O^{t-2}, O^{t-1}]$; \\
Merge the predicted value $\hat{N}^{t}$ in the belief state vector $S_\text{blf}^t$; \\
\textbf{ACB}: Select $A_\text{ACB}^t= \pi(S^t;\bm{\theta}_\text{ACB}) + {\cal N}^t$; \\
\textbf{BO}: \lIf{$p^t_{\epsilon}<\epsilon$}{select a random action $A_\text{BO}^t$ from $\cal A_\text{BO}$}   
\hspace{+0.6cm} \lElse{Select $A_\text{BO}^t= \mathop {\text{argmax}}\limits_{{a\in {\cal A}_\text{BO}}} Q(S^t, a; \bm{\theta_\text{BO}})$} 
\textbf{DQ}: \lIf{$p^t_{\epsilon}<\epsilon$}{select a random action $A_\text{DQ}^t$ from $\cal A_\text{DQ}$}   
\hspace{+0.6cm} \lElse{Select $A_\text{DQ}^t= \mathop {\text{argmax}}\limits_{{a\in {\cal A}_\text{DQ}}} Q(S^t, a; \bm{\theta_\text{DQ}})$} 
BS broadcasts action $A^t=\{A^t_\text{ACB}, A^t_\text{BO}, A^t_\text{DQ}\}$, and backlogged IoT devices execute RACH;\\
BS observes $S^{t+1}$, and estimate backlog value $\tilde{N}^t$ as well as calculates $R^{t} = V_{s}^{t+1}$; \\
Storing transition $(S^{t}, \tilde{N}^t)$ in the replay memory of RNN traffic predictor;\\
Storing each transition $(S^{t}, A_\text{ACB}^t, R^{t}, S^{t+1})$, $(S^{t}, A_\text{BO}^t, R^{t}, S^{t+1})$, and $(S^{t}, A_\text{DQ}^t, R^{t}, S^{t+1})$ in each DRL replay memory $M_\text{ACB}$, $M_\text{BO}$, and $M_\text{DQ}$, respectively;\\
Sampling random minibatch of transitions from each replay memory; \\
Calculate the loss of RNN $ L^{t}({\bm{\theta}_\text{RNN}})$ using Eq. \eqref{eq19}, the loss of Q-function for the BO and the DQ schemes $\nabla L(\theta_\text{DQN}^t)$ using Eq. \eqref{eq12}, and the loss of critic $\nabla L(\bm{w}_\text{ACB}^t)$ and actor $\nabla L(\bm{\theta}_\text{ACB}^t)$ for the ACB scheme using Eq. \eqref{eq13} and Eq. \eqref{eq14}, respectively;\\
Perform a gradient descent for the RNN predictor and each primary DRL network; \\
Update the target DRL networks using: $\bm{\bar{w}}^t \leftarrow \sigma\bm{w}^t + (1-\sigma)\bm{\bar{w}}^t $, and $\bm{\bar{\theta}}^t \leftarrow \sigma\bm{\theta}^t + (1-\sigma)\bm{\bar{\theta}}^t $.
}
}
\end{algorithm}

\subsubsection{DRL-based Parameter Configuration}
As seen in Fig. \ref{fig5} (b), the newly predicted traffic value is input into the DRL agents along with the historical traffic values, where this vector of backlog values is treated as the approximate belief state of the DRL agents. Recall that the historical length $T_o$ is manually selected according to the expected memory for time-correlation recognition. To do so, the belief state variable of each agent can be written as $S_\text{blf}^t=[A^{t-T_o},\hat{N}^{t-T_o+1},A^{t-T_o+1},\hat{N}^{t-T_o+2},...,A^{t-1},\hat{N}^{t}]$. Similar as the training of multi-agent DRL given in Sec. \ref{sec5.1}, every DRL agent shares their historical action selection $A=\{A_\text{ACB}, A_\text{BO}, A_\text{DQ}\}$ to enable cooperation. Different from Sec. \ref{sec5.1}, not only the DRL agents are trained, but also the RNN-based traffic predictor is updated after each frame. In the following, the implementation of decoupled learning strategy for multi-agent DRL training is shown in \textbf{Algorithm \ref{a3}}.

In the training process, the evolution of RNN traffic predictor and the DRL agents are mutual correlated, thus they need to be trained in parallel. This parallel adaptation can be implemented in either simulation or practice, while the former case enables a pre-training to ease practical implementation, and the latter case allows all agents to adapt to the realistic traffics. Specifically, the weights of DRL agents are able to be initialized by first training in offline experiments based on available traffic models, which would considerably reduce the time and computational resource needed for their convergence of training in practice. However, this assumed traffic models may mismatch with the practical traffic statistics, thus the adaptation in practice is still necessary. This pre-training can be treated as a case of meta-learning. An numerical example will be given in the next subsection.

\subsection{Numerical Results and Evaluation}\label{sec5.3}

In this subsection, numerical experiments are conducted to evaluate the performance of the hybrid ACB, BO, and DQ schemes in terms of the access success devices number, the access delay, and the energy consumption. Particularly, the following figures represent results of these three KPI taking into account all IoT devices whatever success or fail in access, rather than that in the proposed algorithms, where only the access success IoT devices can be observed in the BS. 
The calculation of access delay and energy consumption is based on the functions given in Eq. \eqref{eq5} and Eq. \eqref{eq6}, and the related parameters are listed in Table \ref{table2} provided by 3GPP for cellular mIoT systems \cite{3GPP2015Cellular}. We adopt the similar traffic profile, network parameters, and hyperparameters setting of neural networks as Sec. \ref{sec4.3}. Importantly, hybrid scheme integrates the features of the ACB, the BO, and the DQ schemes, where they may be jointly executed according to control of the DRL agents.

                        

\begin{table}[htbp!]
\setcounter{table}{0}
	\centering
	\caption{RL Hyperparameters \vspace*{-0.0cm}}
	{\renewcommand{\arraystretch}{0.6}
		\begin{tabular}{|*{1}{p{5cm}}|*{1}{p{2cm}} |*{1}{p{5cm}}|*{1}{p{2cm}} |  }
			\hline
			\rowcolor{Gray}
		   \bf{Parameters}   &    \bf{Value}$\vphantom{\Big(}$  & 
		   \bf{Parameters}   &    \bf{Value}$\vphantom{\Big(}$  \\ \hline
		   $\vphantom{\big(}$Synchronization time $T_\text{sy}$ & 0.65 s & 
		   $\vphantom{\big(}$Msg1 transmission time $T_\text{Msg1}$ & 0.084 s \\
           $\vphantom{\big(}$Msg2 receiving time  $T_\text{Msg2}$ & 0.345 s & 
           $\vphantom{\big(}$Msg3 transmission time $T_\text{Msg3}$ & 0.08 s \\
           $\vphantom{\big(}$Msg4 receiving time  $T_\text{Msg4}$ & 0.345 s & 
           $\vphantom{\big(}$Transmit power $P_\text{Msg1}$, $P_\text{Msg3}$ & 0.545 W \\
           $\vphantom{\big(}$Receive power  $P_\text{sy}$, $P_\text{Msg2}$, $P_\text{Msg4}$ & 0.09 W  &  $\vphantom{\big(}$  & $\vphantom{\big(}$  \\
                        \hline
                        
		\end{tabular}
	}
	\vspace*{-0.1cm}
	\label{table2}
\end{table}

\begin{figure}[htbp!]
    \begin{center}
        \subfigure[\scriptsize The average energy consumption]{
        \includegraphics[width=0.48\textwidth]{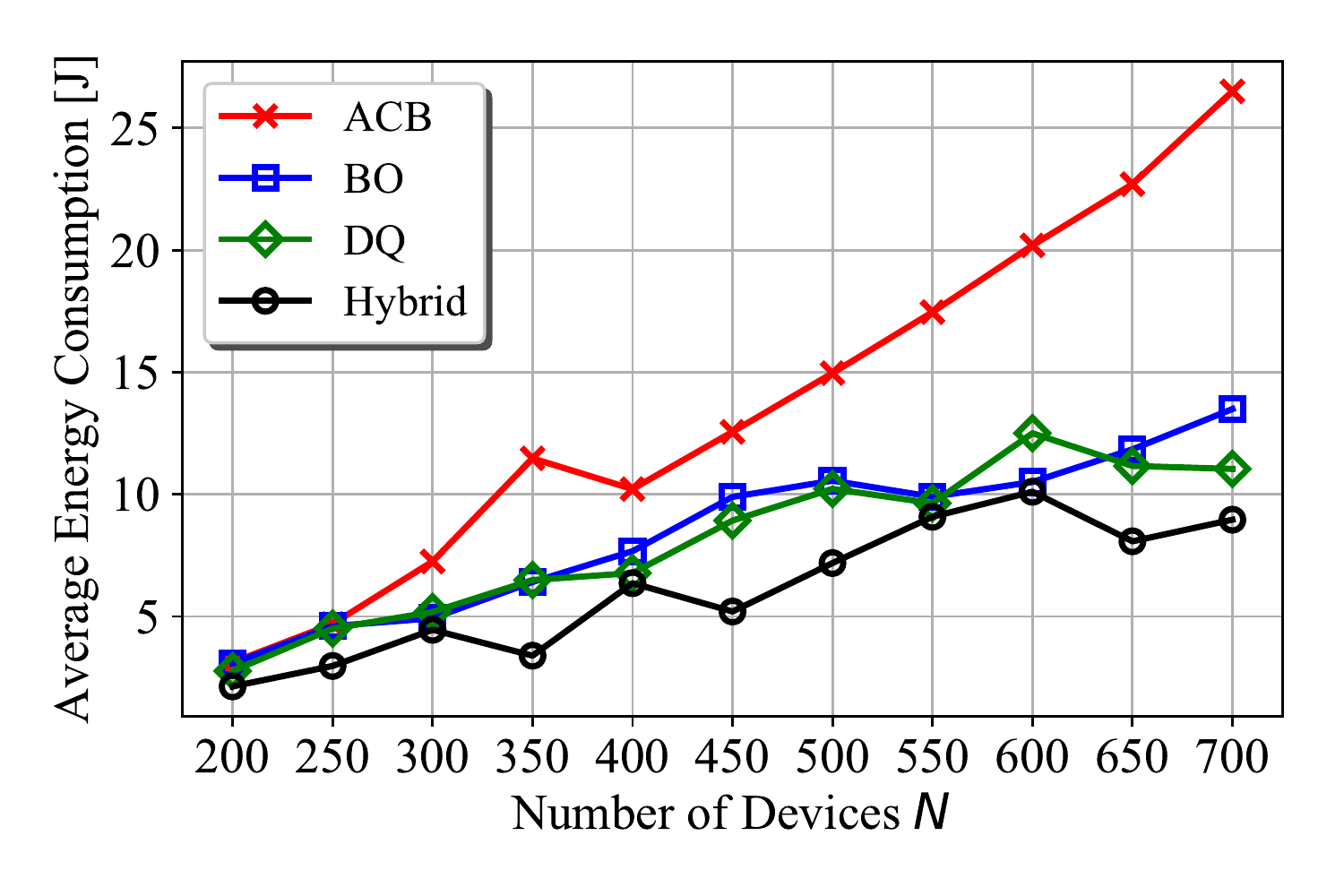}}
        \subfigure[\scriptsize The average access delay]{ \includegraphics[width=0.48\textwidth]{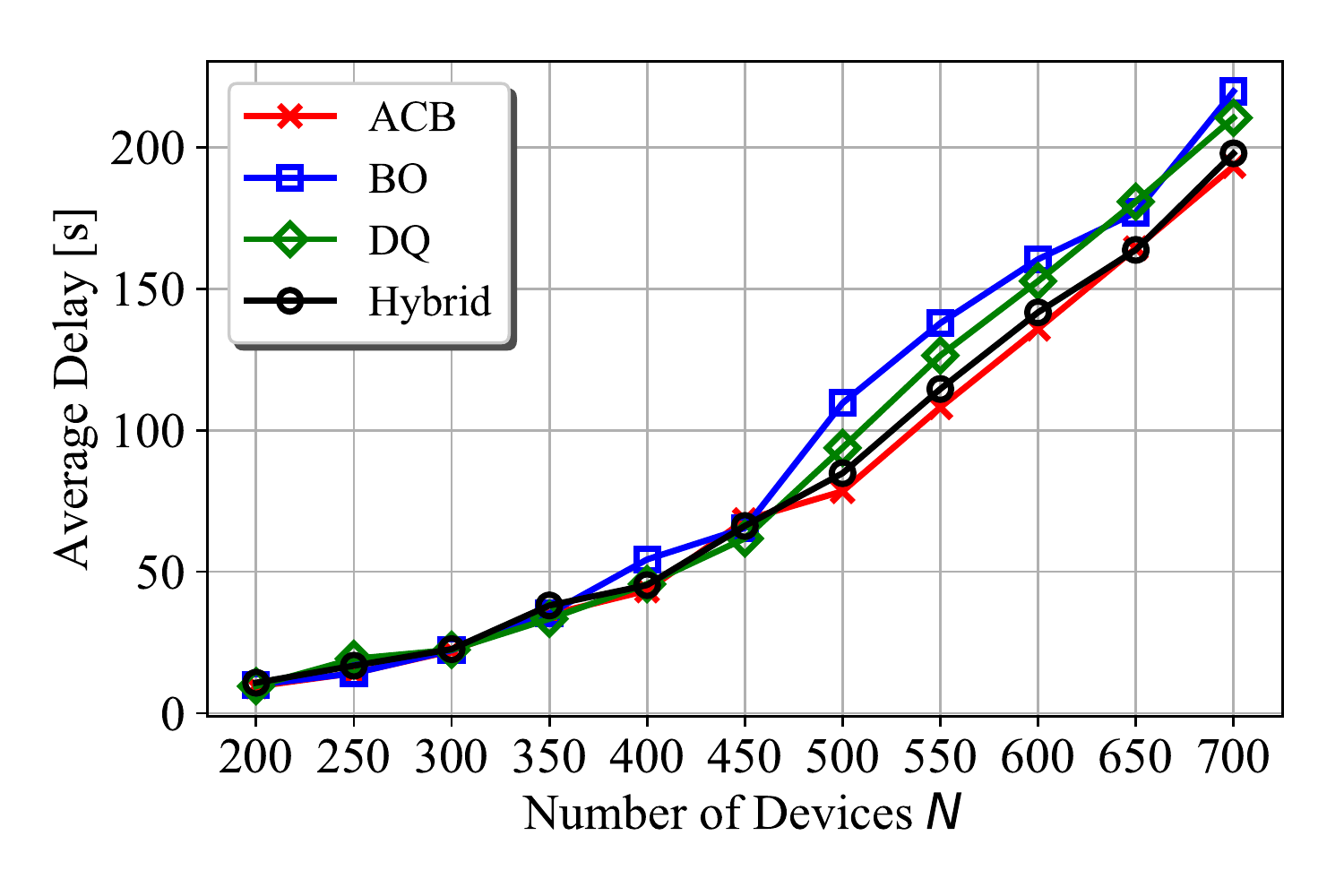}}
         \caption{(a) average energy consumption, and (b) average access delay per IoT device per episode.}
         \vspace*{-0.8cm}
         \label{figr4}
    \end{center}
\end{figure}

We start by illustrating the average energy consumption and the average access delay per IoT device per episode of each scheme versus the presence of devices $N$ in Fig. \ref{figr4} (a) and Fig. \ref{figr4} (b), respectively. The hybrid scheme is implemented using the conventional multi-agent cooperative DRL proposed in Sec. \ref{sec5.1}. Note that the results of each point are obtained by evaluating a scheme with its DRL agents that are trained after 10$^5$ episodes. In particular, the DRL agents for each scheme used in Fig. \ref{figr4} (a) and Fig. \ref{figr4} (b) are optimized by using the reward weights with $x_{s}:x_{d}:x_{e}=1:0:1$ and $x_{s}:x_{d}:x_{e}=1:1:0$ in Eq. \eqref{eq2}, respectively. More specifically, the DRL agents in Fig. \ref{figr4} (a) aim at saving the energy of IoT device while optimizing the  number of access success IoT devices with the reward of energy consumption as $x_{e}=1$, while the DRL agents in Fig. \ref{figr4} (b) aim at reducing the access delay while optimizing the number of access success IoT devices with the reward of access delay as $x_{d}=1$. In both Fig. \ref{figr4} (a) and Fig. \ref{figr4} (b), it can be seen that increasing the number of IoT devices $N$ increases both the average energy consumption and the average access delay of all schemes, due to that the increase of traffic statistics leads to more collisions.

\vspace*{-0.0cm}

In Fig. \ref{figr4} (a), the performance of the ACB scheme is notably worse than the other schemes, due to that IoT devices tends to waste more energy when they repeatedly listen to the ACB factor to ask access permissions. In Fig. \ref{figr4} (b), the ACB scheme slightly outperforms the BO and the DQ schemes, when the number of devices $N$ is bigger than $450$. This is due to that the ACB scheme can accurately set the access request probability of each device at every frame, while the BO and the DQ scheme can only schedule re-transmissions into relatively long future frames. Due to this feature, the ACB scheme is more capable in accurately alleviating the overloaded traffic to reduce access delay, but it also sacrifices the energy consumption as shown in Fig. \ref{figr4} (a). It is also observed that the hybrid scheme always outperforms the other single schemes in terms of energy saving in Fig. \ref{figr4} (a), while it is close to, but is no better than, the ACB scheme in terms of delay reduction in Fig. \ref{figr4} (b). The former phenomenon is due to that the hybrid scheme is capable of adjusting to the complex energy saving requirement by optimally balancing all three schemes, while the latter phenomenon is due to that the hybrid scheme is learned to optimize the performance by mostly relying on the ACB control, which is the best method to reduce access delay.

\begin{figure}[t]
    \begin{center}
        \subfigure[\scriptsize The average success accesses]{
        \includegraphics[width=0.32\textwidth]{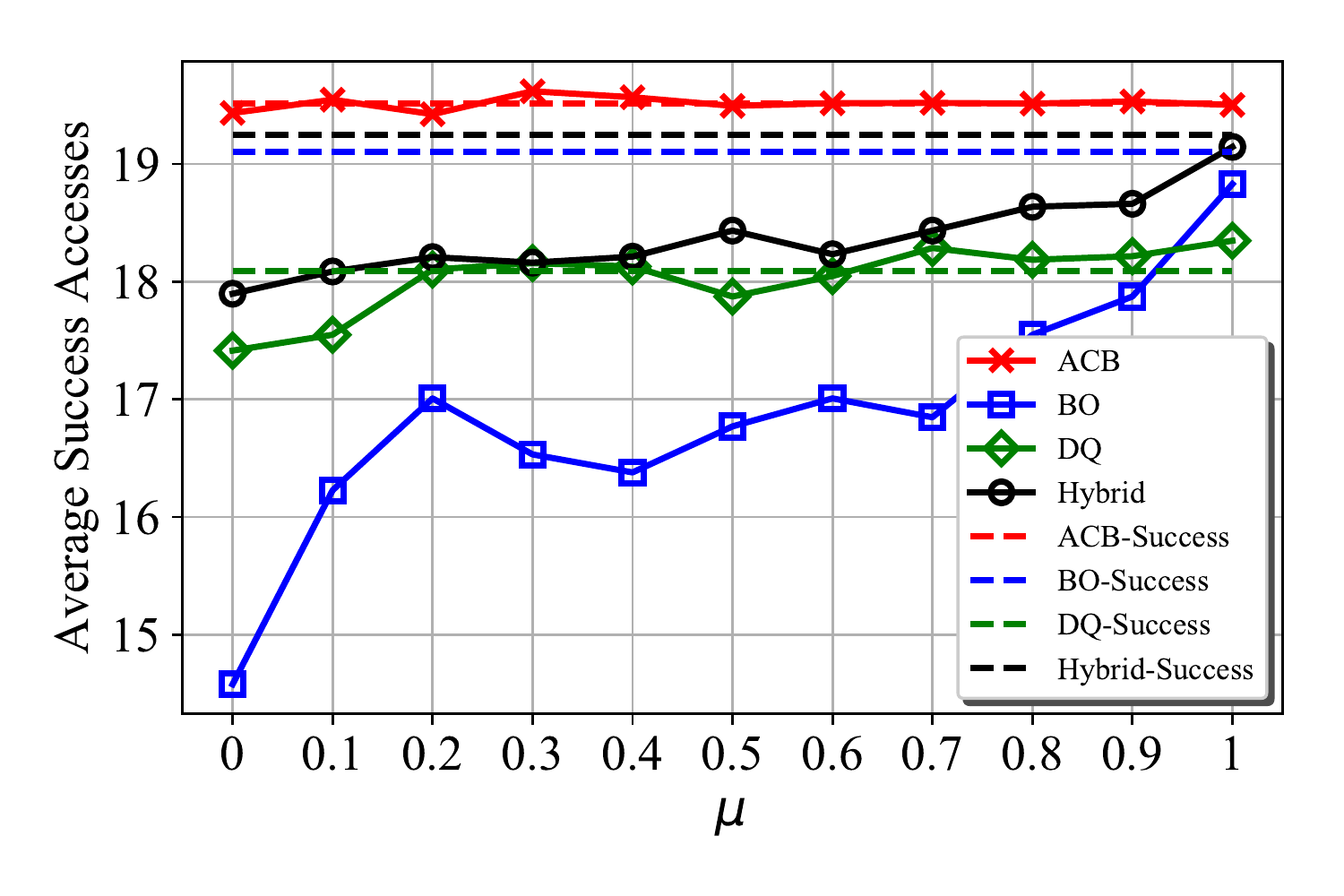}}
        \subfigure[\scriptsize The average energy consumption]{
        \includegraphics[width=0.32\textwidth]{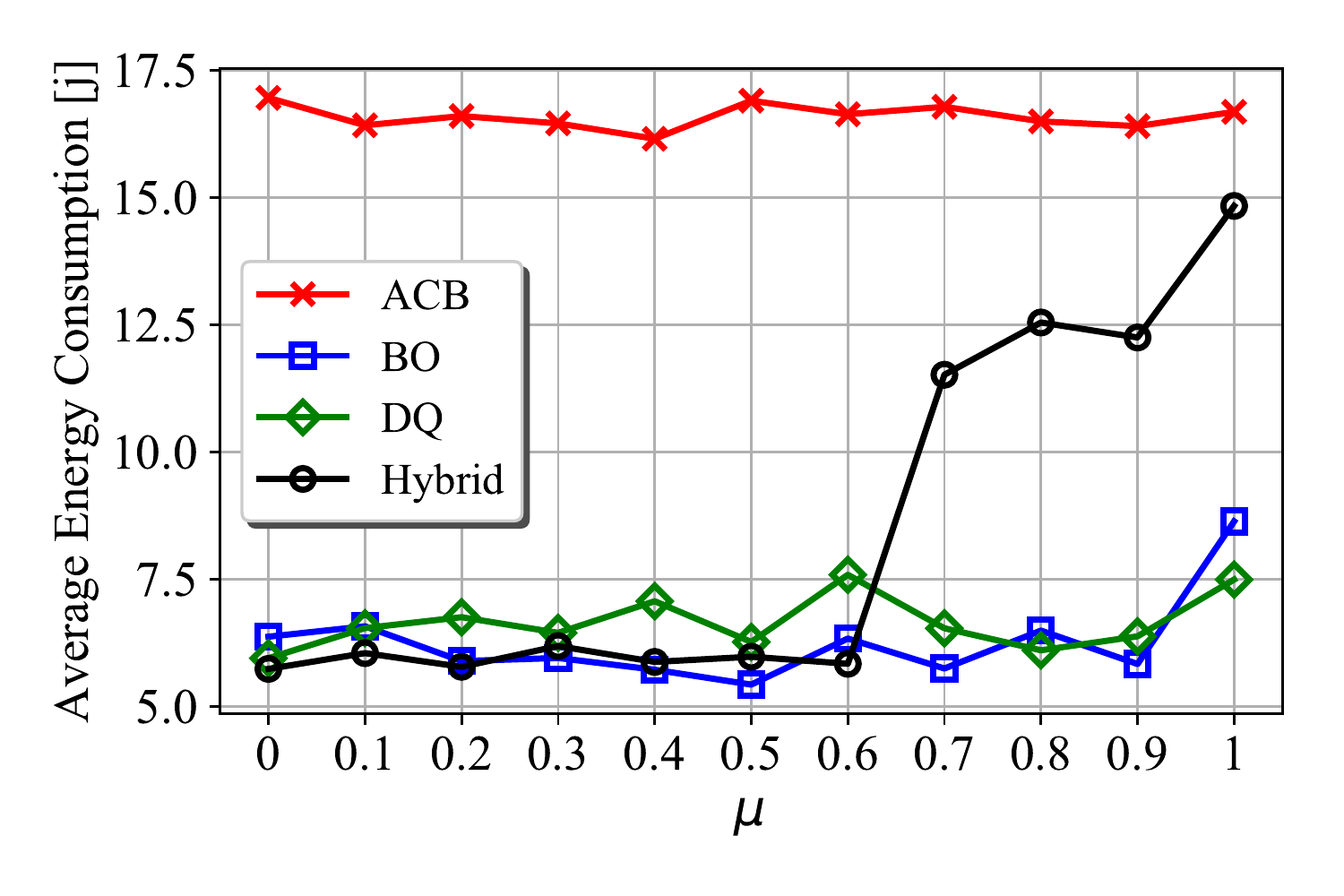}}
        \subfigure[\scriptsize The average access delay]{ \includegraphics[width=0.32\textwidth]{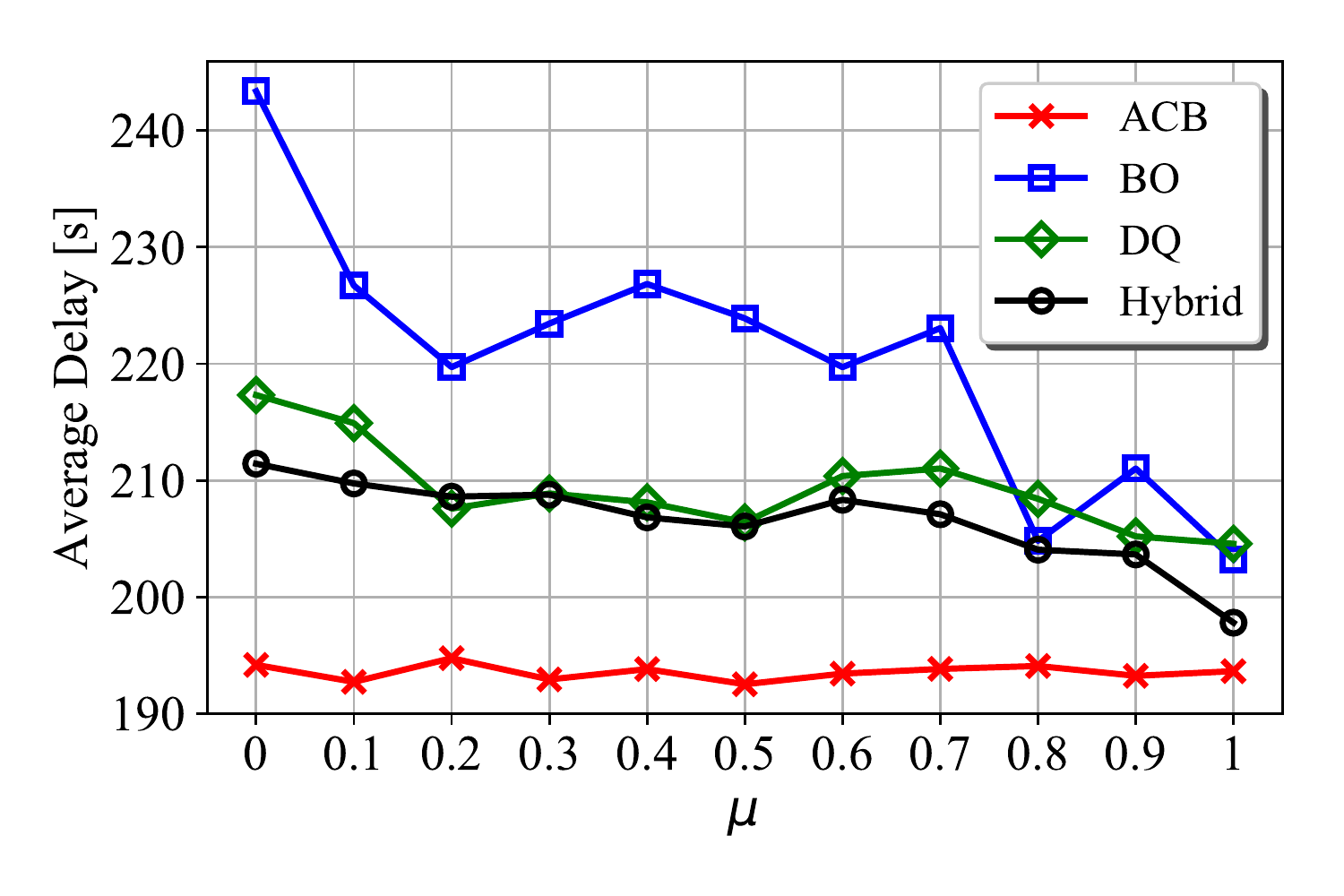}}
         \caption{The (a) average success accesses, (b) average energy consumption, and (c) average access delay per IoT device per episode.}
         \vspace*{-0.8cm}
         \label{figr5}
    \end{center}
\end{figure}

We now consider the reward weights are defined by $x_{s}:x_{d}:x_{e}=1:\mu:1-\mu$, where $\mu \in [0,1]$. By selecting different value of $\mu$, the DRL agents are able to optimize the system with different priorities between energy and delay. In the following, we plot the average number of access success devices, the average energy consumption, and the average access delay per IoT device per episode of each scheme versus the priority value $\mu$ in Fig. \ref{figr5} (a), Fig. \ref{figr5} (b), and Fig. \ref{figr5} (c), respectively. In Fig. \ref{figr5} (a), ``ACB-Success", ``BO-Success", ``DQ-Success", and ``Hybrid-Success" with dashed lines represent these four schemes trained by using the reward only considering the number of access success devices (reward weights $x_{s}:x_{d}:x_{e}=1:0:0$). We observe that the average number of access success devices of the ACB scheme is always longer than that of the other schemes, due to that the ACB scheme can directly determine the access request probability of each IoT device in each frame. Except from the ACB scheme, the performance of the other schemes increase with the increase of $\mu$, due to that the access delay is inversely proportional to the number of access success devices. It can be seen that, with the increase of the delay weight ($\mu$), the average energy consumption of the BO, DQ, and hybrid schemes increases in Fig. \ref{figr5} (b), whereas the average access delay of these schemes decreases in Fig. \ref{figr5} (c). This shows that the energy-delay trade-off of the network system can be flexibly balanced by selecting different weights in the proposed hybrid reward. When $\mu$ is smaller than 0.3, we observe that the proposed hybrid scheme sacrifices the delay property, and consumes the minimum energy. We further observe that, with the increase of $\mu$, the average access delay of the proposed hybrid scheme gradually closes to that of the optimal ACB scheme, and it always outperforms the BO and DQ schemes. This demonstrates the capability of the proposed hybrid scheme to flexibly adapt to different performance requirements of the network.

\vspace*{-0.0cm}

\begin{figure}[t!]
    \begin{center}
        \includegraphics[width=0.5\textwidth]{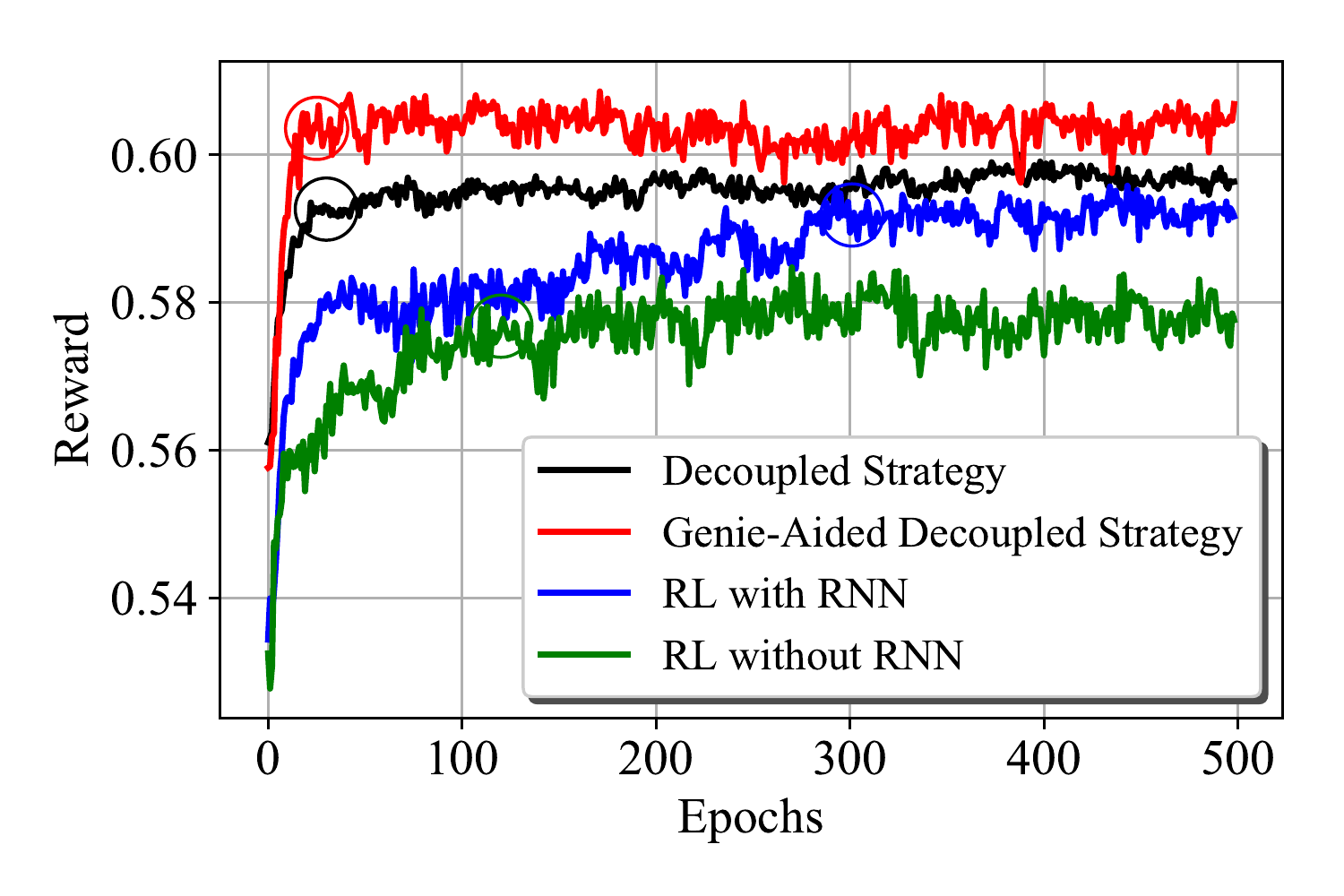}
        \vspace*{-0.5cm}
         \caption{
          Average reward per frame as a function of each epoch in the online adaptation phase
          }
          \vspace*{-1cm}
         \label{figr6}
    \end{center}
\end{figure}

Fig. \ref{figr6} plots the evolution (averaged over 100 training trails) of the average reward per frame as a function in the online phase for the proposed decoupled strategy, genie-aided decoupled strategy, as well as the conventional RL methods with and without the help of RNN. The proposed decoupled strategy is conducted according to Algorithm \ref{a3}, while genie-aided decoupled strategy uses the same algorithm, except that the training is based on the criterion Eq. \eqref{eq19} with the ideal label $N^{t}$ in lieu of the estimated label $\tilde{N}^{t}$. The conventional RL methods are based on the method given in Sec. \ref{sec4.2} and Sec. \ref{sec5.1}, and specifically, the one without RNN uses fully-connected neural network with two hidden layers, each with 128 ReLU units. For fairly considering both energy consumption and access delay, the reward weights are defined by $x_{s}:x_{d}:x_{e}=2:1:1$ ($\mu =0.5$). The approximated converging point of each scheme is highlighted by circles. We observe that the conventional RL without RNN has the worst performance, due to that its fully-connected neural network cannot capture the time correlation conducted by time-varied traffics and devices' queuing processes. This showcases the necessity of using RNN in RACH procedure optimization.

\vspace*{-0.0cm}

We then observe that our proposed decoupled strategy can quickly adapt to the network conditions, where its training speed is about $10$ times than that of RL with RNN, and its obtained average reward is also much higher than that of RL with RNN. It is also seen that both these performance of the decoupled strategy is very close to that of the ideal genie-aided one. This demonstrates that the proposed decoupled strategy is capable to optimize the RACH schemes with better performance and faster converging speed.

\subsection{A Case Study of NarrowBand IoT Networks}\label{sec5.4}

To show the effectiveness of our proposed decoupled strategy, we now consider a more practical NarrowBand (NB)-IoT scenario according to the system model provided in our prior work \cite{jiang2019deep,jiang2019cooperative}, in which an NB-IoT network composed of an evolved Node B (eNB) and $N = 30000$ static IoT devices with time-limited Beta traffic profile. The eNB supports three Coverage Enhancement (CE) groups to provide access for IoT devices with different location. Each IoT device determines their CE identity according to the their distance to the associated eNB. After that, each IoT device executes RACH as well as uplink data transmission according to the received system information that relates to their CE identity. More simulation details can be found in our prior work \cite{jiang2019deep,jiang2019cooperative}, and the 3GPP reports \cite{3GPP2015Cellular,3GPP2017MAC}.

Different from the results in Sec. \ref{sec5.3} independently focusing on RACH procedure, this model considers an IoT device that is successfully served only when it succeeds in both RACH and uplink data transmission. In each frame, the eNB allocates the radio resources to accommodate the RACH procedure for each CE group with the remaining resources used for uplink data transmission. Considering the target of maximizing the number of served IoT devices, the challenge is to optimally balance the allocations of channel resources between the RACH procedure and data transmission, as well as among each CE group. Similar as \cite{jiang2019deep}, we assume the eNB can flexibly select the parameters of the number of RACH periods  $n^t_{\text{Rach},i}$, the number of available preambles $f^t_{\text{Prea},i}$, and the repetition value $n^t_{\text{Repe},i}$ in each group $i$ at each frame $t$. Extended from \cite{jiang2019deep}, we further consider that the eNB can flexibly select the ABC and the BO factors to alleviate traffic overload. The DQ scheme is not standardized in NB-IoT networks \cite{3GPP2017MAC}, so as it is not used in this case.

\vspace*{-0.0cm}
\begin{figure}[t!]
    \begin{center}
    \begin{minipage}[t]{0.47\textwidth}
        \includegraphics[width=1\textwidth]{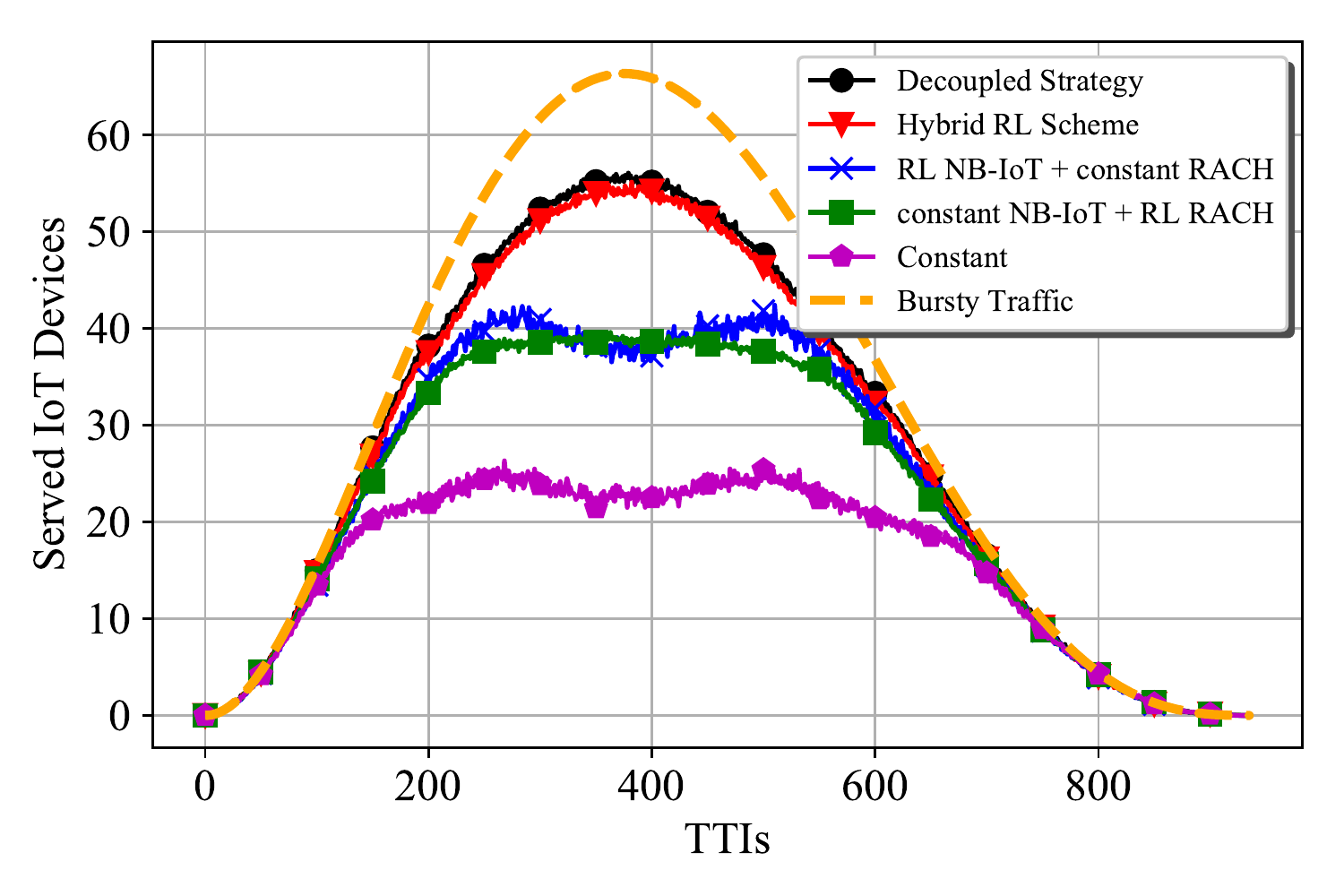}
         \caption{
          The average number of successfully served IoT devices per frame during one bursty traffic duration. The dashed line represents the average number of generated packets per frame.
          }
          \vspace*{-1.1cm}
         \label{figr7}
        \end{minipage}
        \hspace*{+0.5cm}
      \begin{minipage}[t]{0.47\textwidth}
        \includegraphics[width=1\textwidth]{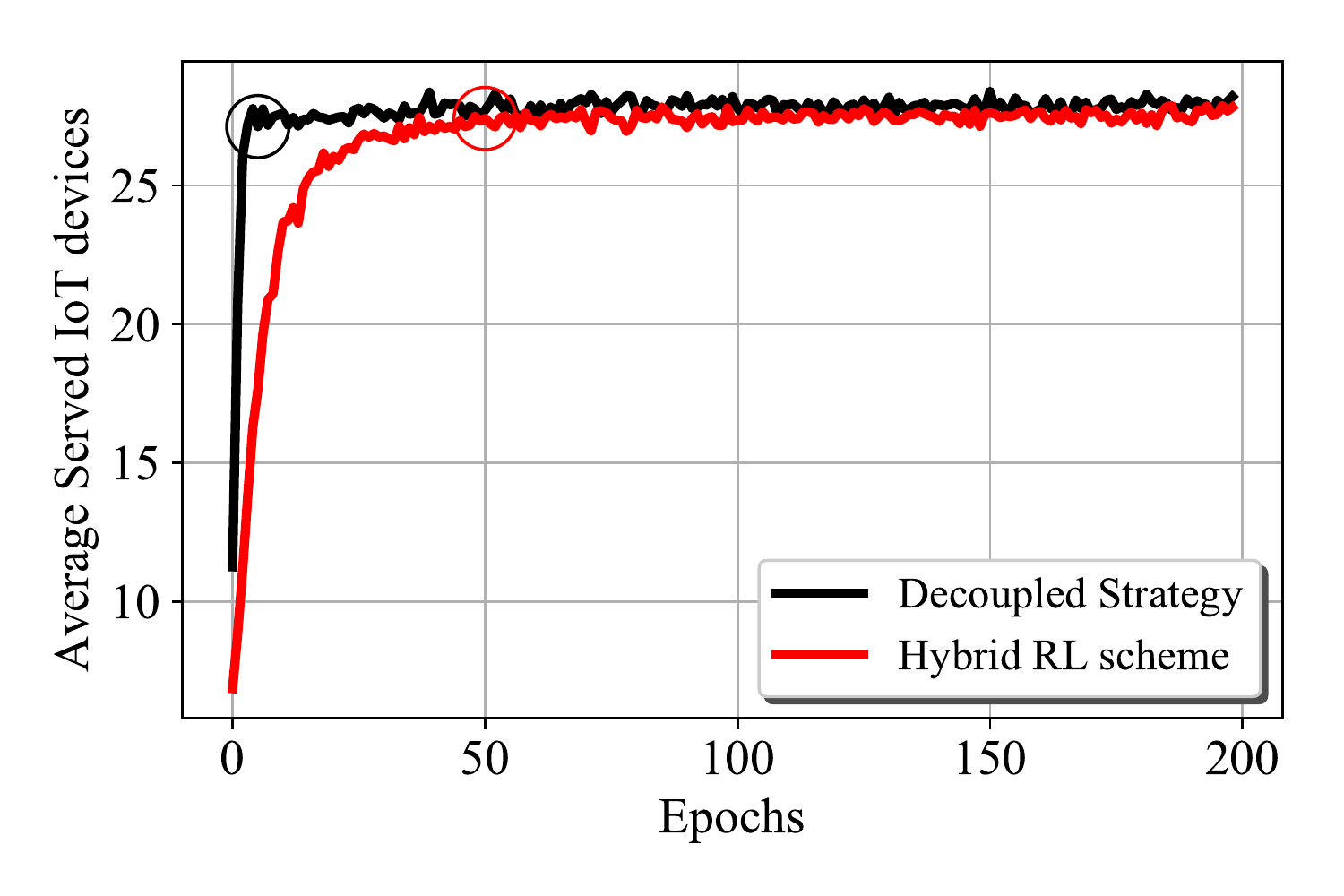}
         \caption{
          The average number of successfully served IoT devices per epoch during training.
          }
          \vspace*{-1.1cm}
         \label{figr8}
        \end{minipage}
    \end{center}
\end{figure}

Fig. \ref{figr7} compares the number of successfully served IoT devices per frame during one epoch. The result of each curve is averaged over 1000 testing epochs. The average number of newly generated packets conducted by the time limited Beta profile is shown as dashed line. We compare the performance in terms of the average number of successfully served IoT devices among the following five schemes: 1) ``Decoupled Strategy" proposed in Sec. \ref{sec5.2}; 2) ``hybrid RL scheme" given in \ref{sec5.1}; 3) ``RL NB-IoT $+$ constant RACH", which uses hybrid RL scheme to configure NB-IoT factors $\{n^t_{\text{Rach},i}, f^t_{\text{Prea},i}, n^t_{\text{Repe},i}\}$ as well as sets constant BO and ACB factors $\{\beta^t_\text{ACB}, \beta^t_\text{BO}\} = \{0.5, 2\}$ for each CE group; 4) ``Constant NB-IoT $+$ RL RACH", which sets NB-IoT factors $\{[n^t_{\text{Rach},1},n^t_{\text{Rach},2},n^t_{\text{Rach},3}]$, $[f^t_{\text{Prea},1},f^t_{\text{Prea},2},f^t_{\text{Prea},3}]$, $
[n^t_{\text{Repe},1}, n^t_{\text{Repe},2}, n^t_{\text{Repe},3}]\}$=$\{ [4, 2, 1]$, $[48, 36, 24]$, $[2, 16, 32]\}$ as well as configures BO and ACB factors by using hybrid RL scheme; and 5) ``Constant" scheme sets constant parameters during whole epoch including NB-IoT factors $\{[n^t_{\text{Rach},1},n^t_{\text{Rach},2},n^t_{\text{Rach},3}], [f^t_{\text{Prea},1},f^t_{\text{Prea},2},f^t_{\text{Prea},3}], $
$[n^t_{\text{Repe},1}, n^t_{\text{Repe},2}, n^t_{\text{Repe},3}]\}=\{ [4, 2, 1]$ ,$[48, 36, 24]$, $[2, 16, 32]\}$ as well as BO and ACB factors $\{\beta^t_\text{ACB}, \beta^t_\text{BO}\} = \{0.5, 2\}$ for each CE group.

In Fig. \ref{figr7}, we observe that the number of served IoT devices of all schemes are close in the light traffic regions at the beginning and end of the epoch, however, in the period of heavy traffic in the middle of the epoch, that number of served IoT devices follows the order ``Decoupled Strategy"$\approx$`hybrid RL scheme"$>$``Constant NB-IoT $+$ RL RACH"$\approx$``RL NB-IoT $+$ constant RACH"$>$``Constant". This showcases that the more dynamically configured system factors are, the better performance can be achieved. This is due to that the constant factors cannot fit all the traffic statistics, while RL algorithms can always find a global optimal combination of the system parameters to adapt to any online traffic statistics. It is also observed that ``Decoupled Strategy" slightly outperforms `hybrid RL scheme", which demonstrates the capability of the proposed decoupled strategy can better manage the channel resources, and achieve better system performance in the presence of heavy traffic. Fig. \ref{figr8} plots the evolution (averaged over 20 training trails) of the average number of served IoT devices per epoch as a function during the online training for ``Decoupled Strategy" and ``hybrid RL scheme". We observe that our proposed decoupled strategy can quickly adapt to the network conditions, which is about 10 times faster than the conventional hybrid RL scheme.

\section{Conclusion}\label{sec7}

In this paper, we developed a decoupled learning strategy to maximize a multi-objective function that is composed of the number of access success devices, the average energy consumption, and the average access delay. The proposed algorithm jointly optimized multiple RACH schemes, including the ACB, BO, and DQ schemes. Our proposed strategy decoupled the traffic prediction and the parameter configuration. The former predictor uses a GRU RNN model to predict the real-time traffic values of the network system, which captures temporal correlations due to communication mechanisms and irregular traffic generation. The latter controller configures system parameters of each RACH scheme by using multiple DRL agents, where DQN is used to handle discrete action selection for the BO as well as the DQ schemes, and DDPG is used to handle continuous action selection for the ACB scheme. Numerical results shed light on that the RACH schemes can be effectively optimized in a joint manner by using the cooperative training to adapt to any performance requirement, where it outperforms each single RACH scheme. More importantly, by using our proposed decoupled learning strategy, the training speed of the cooperative model accelerates around $10$ times than that of the conventional strategy, it considerably outperforms the conventional strategies. As most of the current schemes on RACH procedure only relies on the central control in the BS, a promising future direction is to develop the learning algorithm that performs the RACH control in a cooperative manner between IoT devices and BS.

\bibliographystyle{IEEEtran}
\bibliography{IEEEabrv,RA_bib}

\end{document}